\documentclass[12pt,a4paper]{article}

\usepackage{jheppub}
\usepackage{amsmath,amssymb,mathtools,bm}
\usepackage{booktabs}
\usepackage{microtype}

\providecommand{\dd}{\mathrm d}
\providecommand{\ii}{\mathrm i}
\providecommand{\ee}{\mathrm e}
\providecommand{\cA}{\mathcal A}
\providecommand{\cF}{\mathcal F}
\providecommand{\cN}{\mathcal N}
\providecommand{\cO}{\mathcal O}
\providecommand{\cZ}{\mathcal Z}

\providecommand{\Es}{E_s}
\providecommand{\HH}{\mathbb H}
\providecommand{\CC}{\mathbb C}
\providecommand{\ZZ}{\mathbb Z}
\providecommand{\diag}{\mathrm{diag}}
\providecommand{\off}{\mathrm{off}}

\title{Lorentzian Regularization of the\\
Type~IIB Superstring Torus Vacuum}
\author[1,2,3]{Thomas Junkai Wang}
\affiliation[1]{Tsinghua University, Beijing 100084, China}
\affiliation[2]{TsingyuAI Corporation, Beijing 100084, China}
\affiliation[3]{Nanjing University, Nanjing 210023, China}
\emailAdd{WangTheoPhys@outlook.com}

\abstract{
We study the one-loop torus vacuum of Type IIB Superstring theory through sector-resolved modular integrals. Building on the \(i\varepsilon\)-prescription and the \(E_s\)-regularized modular-integral framework of Manschot and Wang~\cite{ManschotWang2024}, we construct regularized sector functionals for the closed oriented torus before the final GSO projection. The construction keeps the unprojected spin-sector data explicit and fixes the compact-domain and cusp contributions within a single modular prescription. We also independently cross-check the result with the Lorentzian-inversion reconstruction of modular integrals by Baccianti et al.~\cite{BacciantiChandraEberhardtHartmanMizera2025}. This provides a first direct regularized construction of the unprojected sectors of the Type IIB Superstring torus vacuum.
}


\begin{document}
\maketitle
\flushbottom

\section{Introduction}
\label{sec:introduction}

One-loop amplitudes occupy a distinguished place in perturbative string theory because they are the first amplitudes whose definition depends simultaneously on modular geometry, on the global structure of the integration contour, and on the spectrum of intermediate string states. At genus one, the worldsheet modulus \(\tau=x+\ii y\) is integrated over the modular fundamental domain \(\cF\), giving a modular integral over a non-compact orbifold. The non-compact cusp \(y\to\infty\) is the non-separating degeneration of the torus, where a long tube develops and the worldsheet description approaches propagation over a large proper time. Standard references on genus-one string perturbation theory therefore frame one-loop amplitudes as modular integrals whose analytic content is inseparable from the geometry of the cusp itself~\cite{GreenSchwarzWitten1988,Polchinski1998Vol1,Polchinski1998Vol2,DHokerPhong1988Geometry}.

For non-vacuum closed-string amplitudes, the modular integral is accompanied by integrations over puncture positions on the torus, and these puncture integrals generate the Koba--Nielsen factors that carry the kinematic dependence of the external states. In the low-energy expansion of genus-one type II amplitudes, this structure is reorganized in terms of modular graph functions and their generalizations, which capture the modular dependence left after the puncture integrations have been expanded order by order in momenta~\cite{GreenRussoVanhove2008,DHokerGreenVanhove2015ModularStructure,DHokerGreenGurdoganVanhove2017,Zerbini2016SingleValued,DHokerKaidi2016Hierarchy,Basu2019Eigenvalue}. The vacuum problem isolates the modular integral and its cusp regularization before the additional kinematic structure from punctures and Koba--Nielsen factors is introduced.

The Euclidean torus integral by itself does not completely specify the Lorentzian amplitude once degeneration regions allow intermediate string states to go on shell. In that regime one must choose an \(i\varepsilon\) prescription in moduli space, just as one chooses a Feynman prescription in field theory. Witten formulated this problem directly in string perturbation theory by showing how the physical amplitude is defined by a Lorentzian continuation of the integration cycle in complexified moduli space~\cite{Witten2013IEpsilon}. Related string-field-theoretic treatments of off-shell amplitudes, integration cycles, and picture-changing subtleties provide another formulation of the same need to define superstring amplitudes globally rather than only by a local Euclidean integrand~\cite{Zwiebach1993ClosedSFT,Sen2015Offshell,SenWitten2015PCO}. From the worldsheet point of view, the same degeneration data control discontinuities and unitarity cuts: the long-tube region is where the modular integral resolves into on-shell propagation, and recent work has made this relation explicit for one-loop string amplitudes through worldsheet-cut constructions~\cite{EberhardtMizera2022Cuts,BanerjeeEberhardtMizera2025}.

Modern evaluations of one-loop string amplitudes have consequently developed along several complementary lines. Worldsheet cuts expose the physical discontinuity structure and identify the states that propagate through degeneration channels~\cite{EberhardtMizera2022Cuts,EberhardtMizera2023Open}. Lorentzian inversion and Rademacher-contour methods reorganize modular integrals cusp by cusp and provide convergent representations adapted to analytic continuation~\cite{BacciantiChandraEberhardtHartmanMizera2025}. In that setting, Rademacher contours give a global all-cusp organization of modular-integral data, while a single Lorentzian long-tube contour gives the local standard-cusp building block. At the same time, type II four-graviton amplitudes and their protected low-energy couplings have long served as probes of genus-one modular integration, duality constraints, and threshold behavior~\cite{GreenGutperle1997DInstanton,GreenGutperleVanhove1997OneLoop11D,GreenSethi1998Supersymmetry}; the exact finite-\(\alpha'\) four-graviton amplitude is now a sharper setting for seeing how these analytic tools control a fully kinematic one-loop observable~\cite{BacciantiEberhardtMizera2025b}. The regularized modular-integral prescription of Manschot and Wang rewrites one-loop modular integrals in terms of Fourier data and generalized exponential integrals, making the cusp contribution explicit mode by mode~\cite{ManschotWang2024,WangThesis2025}. The present paper applies this prescription in the simplest closed-string setting where the sector data can be organized completely and compared with exact cancellation identities.

This setting is the type IIB genus-one torus vacuum.  In the usual
\(SO(8)\)-character language, Jacobi's abstruse identity equates the vector
and spinor characters in the signed type IIB torus integrand
~\cite{GreenSchwarzWitten1987,GreenSchwarzWitten1988,Polchinski1998Vol2}.  The
regularized construction acts one step earlier: the auxiliary spin-structure
sectors of the oriented closed torus are separated, the modular-integral map is
applied to each sector, and the physical GSO projection is imposed at the end.
The Lorentzian and \(E_s\)-regularized prescriptions are compared on these
sector functionals.

The broader genus-one problem is represented by the closed-string amplitude
with \(n\) insertions,
\begin{equation}
\cA_{1,n}
=
\int_{\cF}\dd\tau\wedge \dd\bar\tau
\int_{T^2_\tau}\prod_{i=1}^{n}\dd z_i\wedge \dd\bar z_i\,
\langle V_1\cdots V_n\rangle .
\label{eq:intro-A1n}
\end{equation}
This formula exhibits the two integrations that appear at one loop: the modular integral over the torus modulus and, in the non-vacuum case, the puncture integrals over the torus itself. The present paper specializes to the vacuum problem, for which the puncture integrations are absent and one is left with a modular integral of the form
\begin{equation}
\cA_{1,0}
=
\int_{\cF}\dd\tau\wedge \dd\bar\tau\,
y^{-s}f(\tau,\bar\tau).
\label{eq:intro-A10}
\end{equation}
The advantage of the vacuum specialization is that it isolates the modular regularization step from the extra kinematic structure carried by punctures and Koba--Nielsen factors.

The regularization map is most naturally organized after a Fourier expansion of the modular integrand,
\begin{equation}
f(\tau,\bar\tau)
=
\sum_{m,n}F(m,n)q^m\bar q^n,
\qquad
q=\ee^{2\pi \ii \tau}.
\label{eq:intro-Fourierf}
\end{equation}
In the regularized modular-integral prescription of
Manschot and Wang, the integral is organized as the mode sum
\begin{equation}
I^r[f]
=
\sum_{m,n}F(m,n)L^r_{m,n,s}.
\label{eq:intro-Irf}
\end{equation}
Only the mode-sum form is needed at this stage. The modular integral is
reorganized into Fourier data and regularized mode blocks, while the explicit
definition of \(L^r_{m,n,s}\), including the compact-domain part and the cusp
subtraction, belongs to the detailed construction of the regularized mode
integrals.

For the type IIB torus vacuum we study
\begin{equation}
Z^{\rm IIB}_{0,1}
=
\ii \cN
\int_{\cF}
\frac{\dd\tau\wedge \dd\bar\tau}{y^6}\,
\cZ_{\rm IIB}(\tau,\bar\tau),
\label{eq:intro-ZIIB01}
\end{equation}
where the factor \(y^{-6}\) fixes the modular weight parameter to \(s=6\) in the later regularized mode blocks. The full type IIB integrand is decomposed as
\begin{equation}
\cZ_{\rm IIB}
=
\cZ_{VV}
-\cZ_{VS}
-\cZ_{SV}
+\cZ_{SS}.
\label{eq:intro-ZIIBsectors}
\end{equation}
The four functions \(\cZ_{XY}\) are the auxiliary sector integrands on which
the regularized modular-integral map acts before the final physical
GSO contraction is taken.  The paper is concerned with the four sector
functionals \(I^r_{XY}\), not with re-deriving the final signed zero.

Each auxiliary sector carries its own Fourier expansion,
\begin{equation}
\cZ_{XY}(\tau,\bar\tau)
=
\sum_{m,n}F_{XY}(m,n)q^m\bar q^n.
\label{eq:intro-ZXYFourier}
\end{equation}
The regularized sector integral is
\begin{equation}
I^r_{XY}
=
\sum_{m,n\ge0}F_{XY}(m,n)L^r_{m,n,6},
\qquad
\cA_{XY}=\ii I^r_{XY}.
\label{eq:intro-sector-functional}
\end{equation}
This is the main object of the paper.  The Type IIB specialization is not a
literal copy of the Type I setting: the worldsheet is a single oriented closed
torus, the coefficients are closed left-right products, the vacuum measure
fixes \(s=6\), and the sector expansion is non-polar.  The coefficient input
also admits the cusp-theoretic description developed below from the
\(\Gamma(2)\) theta-block orbit and the Rademacher coefficient construction
~\cite{ChengDuncan2012,BacciantiChandraEberhardtHartmanMizera2025}.  The
constant mode, the finite all-mode sector value, and the final GSO projection
are evaluated only after the regularized block and the sector coefficients
have been fixed.
After the sector functionals have been defined, the Jacobi identity supplies
the coefficient-level GSO relation
\begin{equation}
F_{VV}(m,n)-F_{VS}(m,n)-F_{SV}(m,n)+F_{SS}(m,n)=0,
\label{eq:intro-GSOcoeff}
\end{equation}
and hence, by linearity,
\begin{equation}
I^r_{\rm IIB}
=
I^r_{VV}-I^r_{VS}-I^r_{SV}+I^r_{SS}
=0.
\label{eq:intro-IrIIBzero}
\end{equation}
This final cancellation is the signed projection of the sector construction.

The rest of the paper fixes the Type IIB torus amplitude, separates the cusp,
implements the Lorentzian prescription, defines the regularized mode blocks,
assembles the sector amplitudes, evaluates the constant mode, applies the
physical GSO projection, and discusses the relation of the vacuum construction
to the non-vacuum cusp framework. The finite all-mode sector value is collected
in Appendix~\ref{app:numerical-details}.

\section{The type IIB torus amplitude}
\label{sec:type-iib-torus-amplitude}

The closed-string input for the regularized calculation is the genus-one type
IIB torus vacuum written before the GSO cancellation has been contracted. The
vacuum integral itself is standard, but its sector data must be fixed with some
care because the later regularized sums act directly on their Fourier
coefficients. We therefore fix the torus measure, the spin-structure sum, the
\(SO(8)\)-character rewriting, and the auxiliary sector blocks before the
final GSO projection in a single convention.

\subsection{The torus modulus and the genus-one measure}
\label{subsec:torus-measure}

We write the complex structure modulus of the torus as
\begin{equation}
\tau=x+\ii y,
\qquad
q=\ee^{2\pi \ii\tau},
\qquad
\bar q=\ee^{-2\pi \ii\bar\tau},
\label{eq:tauq}
\end{equation}
with \(y>0\). The modular parameter is integrated over the standard
fundamental domain
\begin{equation}
\cF=
\left\{
\tau\in\HH:
-\frac12\le \operatorname{Re}\tau\le \frac12,
\quad
|\tau|\ge 1
\right\},
\label{eq:fundamentaldomain}
\end{equation}
which represents the moduli space \(\HH/SL(2,\mathbb Z)\) of oriented genus-one
worldsheets~\cite{GreenSchwarzWitten1988,Polchinski1998Vol2}. The non-compact
cusp at \(y\to\infty\) will later be the region in which the Lorentzian
prescription and the regularized modular-integral map become relevant, but at
the present stage it enters only through the standard measure.

With an overall normalization denoted by \(\cN\), the type IIB torus vacuum is
written as
\begin{equation}
Z^{\rm IIB}_{0,1}
=
\ii \cN
\int_{\cF}
\frac{\dd\tau\wedge \dd\bar\tau}{y^6}\,
\cZ_{\rm IIB}(\tau,\bar\tau).
\label{eq:IIBvacuum}
\end{equation}
At the operator level, \eqref{eq:IIBvacuum} is the modular integral of a torus
trace over the closed-string Hilbert space. Suppressing the standard ghost and
gauge-fixing factors, the worldsheet trace has the form
\begin{equation}
\operatorname{Tr}_{\mathcal H}\!\left(
q^{L_0-c/24}\bar q^{\bar L_0-c/24}
\right),
\label{eq:torustrace}
\end{equation}
with the spin-structure sum and the \(SO(8)\)-character basis providing a
convenient packaging of that trace in the supersymmetric vacuum sector
~\cite{GreenSchwarzWitten1988,Polchinski1998Vol2}. The factor \(y^{-6}\) in
\eqref{eq:IIBvacuum} is not an arbitrary exponent. In the conventions used
here it combines the modular measure \(y^{-2}\) with the light-cone
non-compact bosonic zero-mode factor \(y^{-4}\), and therefore fixes the
modular-weight parameter that later appears in the regularized mode blocks to
be
\begin{equation}
s=6.
\label{eq:six}
\end{equation}
No regularization formula is introduced here; \eqref{eq:six} fixes only the
input value dictated by the vacuum measure.

\subsection{Spin structures and the type IIB torus integrand}
\label{subsec:spin-structures}

The spin-structure sum fixes the integrand before the character rewriting. In
the standard light-cone gauge conventions, the scalar type IIB torus integrand
is
\begin{equation}
\cZ_{\rm IIB}(\tau,\bar\tau)
=
\frac14
\sum_{\alpha,\beta,\bar\alpha,\bar\beta=0}^{1}
(-1)^{\alpha+\beta+\bar\alpha+\bar\beta}
\frac{
\vartheta_{\alpha\beta}(\tau)^4\,
\vartheta_{\bar\alpha\bar\beta}(\bar\tau)^4
}{
|\eta(\tau)|^{24}
}.
\label{eq:spinstructuresum}
\end{equation}
Here \((\alpha,\beta)\) and \((\bar\alpha,\bar\beta)\) label the left- and
right-moving spin structures, respectively, and \(\eta\) together with the
Jacobi theta constants are normalized as in the standard superstring
references~\cite{GreenSchwarzWitten1987,GreenSchwarzWitten1988,Polchinski1998Vol2}.
Equation \eqref{eq:spinstructuresum} is the oriented closed-string torus
expression. It is not an annulus, M\"obius-strip, or Klein-bottle amplitude,
and it should not be confused with the open or unoriented one-loop geometries
that arise in type I theory~\cite{AngelantonjSagnotti2002,WangThesis2025}.

The spin-structure form is retained because the regularized map acts on sector
Fourier data before the GSO signs are contracted. The full type IIB torus
amplitude is obtained after the spin-structure contraction, whereas the
intermediate sector blocks carry the modular data on which the regularized
mode construction acts.

The standard \(SO(8)\) characters package the non-vanishing theta constants
into the familiar vector and spinor combinations
\begin{equation}
V_8(\tau)=\frac{\vartheta_3(\tau)^4-\vartheta_4(\tau)^4}{2\eta(\tau)^4},
\qquad
S_8(\tau)=\frac{\vartheta_2(\tau)^4}{2\eta(\tau)^4}.
\label{eq:V8S8}
\end{equation}
These obey Jacobi's abstruse identity
\begin{equation}
\vartheta_3(\tau)^4-\vartheta_4(\tau)^4-\vartheta_2(\tau)^4=0,
\label{eq:jacobi}
\end{equation}
and therefore
\begin{equation}
V_8(\tau)=S_8(\tau).
\label{eq:VeqS}
\end{equation}
This identity gives the standard Jacobi cancellation in the full type IIB
torus integrand~\cite{GreenSchwarzWitten1987,GreenSchwarzWitten1988,Polchinski1998Vol2}.
The regularized construction below acts on the auxiliary sector data that appear
before this final cancellation.

Using the character basis, \eqref{eq:spinstructuresum} becomes
\begin{equation}
\cZ_{\rm IIB}(\tau,\bar\tau)
=
\frac{\bigl(V_8(\tau)-S_8(\tau)\bigr)\bigl(\overline{V_8(\tau)}-\overline{S_8(\tau)}\bigr)}
{|\eta(\tau)|^{16}}.
\label{eq:ZIIBcharacter}
\end{equation}
In this basis the left- and right-moving sectors enter symmetrically, and
\eqref{eq:VeqS} fixes the signed integrand pointwise. The sector analysis
below applies the mode-by-mode regularization to the individual sector terms
before the final GSO projection is imposed.

\subsection{Auxiliary sector decomposition}
\label{subsec:sectordecomp}

Expanding the numerator of \eqref{eq:ZIIBcharacter} defines four auxiliary
sector integrands,
\begin{equation}
\cZ_{VV}=\frac{V_8\bar V_8}{|\eta|^{16}},
\qquad
\cZ_{VS}=\frac{V_8\bar S_8}{|\eta|^{16}},
\qquad
\cZ_{SV}=\frac{S_8\bar V_8}{|\eta|^{16}},
\qquad
\cZ_{SS}=\frac{S_8\bar S_8}{|\eta|^{16}},
\label{eq:sectorintegrands}
\end{equation}
where the common arguments \((\tau,\bar\tau)\) are suppressed for brevity.
These are the sector-resolved objects that will later be regularized
individually. They are not separately physical vacuum amplitudes. Rather, they
are auxiliary modular integrands whose final contraction reproduces the
physical type IIB combination
\begin{equation}
\cZ_{\rm IIB}
=
\cZ_{VV}-\cZ_{VS}-\cZ_{SV}+\cZ_{SS}.
\label{eq:GSOintegrand}
\end{equation}
Because \eqref{eq:VeqS} holds identically, \eqref{eq:GSOintegrand} is the
Jacobi-signed contraction of four equal sector blocks. The nontrivial data for
the present calculation are the common sector coefficients carried by those
blocks before the signed contraction is imposed.

\subsection{Holomorphic building blocks and Fourier data}
\label{subsec:fourierdata}

To expose that common data, define the holomorphic building blocks
\begin{equation}
A_V(\tau)=\frac{V_8(\tau)}{\eta(\tau)^8},
\qquad
A_S(\tau)=\frac{S_8(\tau)}{\eta(\tau)^8}.
\label{eq:AVASdef}
\end{equation}
Their Fourier coefficients are introduced through
\begin{equation}
A_X(\tau)=\sum_{m\ge 0}a_X(m)q^m,
\qquad X\in\{V,S\}.
\label{eq:AXFourier}
\end{equation}
Using \eqref{eq:VeqS}, one finds
\begin{equation}
A_V(\tau)=A_S(\tau)=A(\tau),
\label{eq:Acommon}
\end{equation}
with the common closed-string block given by
\begin{equation}
A(\tau)=\frac{\vartheta_2(\tau)^4}{2\eta(\tau)^{12}}.
\label{eq:Atheta}
\end{equation}
The product formulas
\begin{equation}
\eta(\tau)=q^{1/24}\prod_{n\ge 1}(1-q^n),
\qquad
\vartheta_2(\tau)=2q^{1/8}\prod_{n\ge 1}(1-q^n)(1+q^n)^2
\label{eq:etatheta2products}
\end{equation}
therefore imply
\begin{equation}
A(\tau)
=
\frac{\vartheta_2(\tau)^4}{2\eta(\tau)^{12}}
=
8\prod_{n\ge 1}\frac{(1+q^n)^8}{(1-q^n)^8}.
\label{eq:Aproduct}
\end{equation}
This identity already makes the constant term \(8\) and the positivity of the
first few coefficients manifest. Expanding the product then gives the exact
\(q\)-series
\begin{equation}
A(\tau)=8+128q+1152q^2+7680q^3+\cO(q^4).
\label{eq:Aseries}
\end{equation}
In the present project these coefficients are understood as exact outputs of
theta/eta series algebra, not as hand-entered numbers or floating-point fits.

The same coefficient sequence has a cusp-theoretic reconstruction.  Define the
three \(\Gamma(2)\) theta blocks
\begin{equation}
B_i(\tau)=\frac{\vartheta_i(\tau)^4}{2\eta(\tau)^{12}},
\qquad
i=2,3,4.
\label{eq:BiGammaTwoBlocks}
\end{equation}
Then \(A=B_2\).  The \(B_2\) expansion at the cusp \(i\infty\) is non-polar,
as shown in \eqref{eq:Aseries}.  The remaining cusps in the \(\Gamma(2)\)
orbit are represented by the other two theta blocks after the corresponding
cusps are scaled to infinity.  Their width-two local expansions provide the
polar data whose Rademacher coefficient reconstruction gives the positive
coefficients \(d_m=[q^m]B_2\) used below
~\cite{ChengDuncan2012,BacciantiChandraEberhardtHartmanMizera2025}.  This
identification concerns the holomorphic degeneracies only; the regularized
sector functional is obtained later by pairing these coefficients with the
Lorentzian/\(E_6\) mode blocks.

Each auxiliary sector is the left-right product of one holomorphic and one
anti-holomorphic block,
\begin{equation}
\cZ_{XY}(\tau,\bar\tau)
=
A_X(\tau)\,\overline{A_Y(\tau)}
=
\sum_{m,n\ge 0}F_{XY}(m,n)q^m\bar q^n.
\label{eq:ZXYFourier}
\end{equation}
The coefficient matrix factorizes as
\begin{equation}
F_{XY}(m,n)=a_X(m)a_Y(n).
\label{eq:Ffactor}
\end{equation}
Since \eqref{eq:Acommon} implies \(a_V(m)=a_S(m)\) for every \(m\), the four
sector matrices coincide coefficient by coefficient,
\begin{equation}
F_{VV}(m,n)=F_{VS}(m,n)=F_{SV}(m,n)=F_{SS}(m,n).
\label{eq:Fequality}
\end{equation}
This equality is exact and follows from theta-function identities before any
modular integration is attempted.

The constant coefficient is therefore the closed-string product of the left-
and right-moving massless coefficients,
\begin{equation}
F_{XY}(0,0)=8\times 8=64.
\label{eq:F00}
\end{equation}
This is the normalization relevant for the later constant-mode analysis. It
should not be confused with the holomorphic type I Ramond-series normalization
\begin{equation}
\frac{\vartheta_2(\tau)^4}{\eta(\tau)^{12}}=16+256q+\cO(q^2),
\label{eq:typeIRRcomparison}
\end{equation}
which belongs to a single holomorphic block in the open or unoriented setting
and is cited here only as a convention check
~\cite{AngelantonjSagnotti2002,WangThesis2025}. The present paper studies the
closed-string left-right product, so the relevant sector coefficient is
\eqref{eq:F00}, not the holomorphic coefficient in
\eqref{eq:typeIRRcomparison}.

The output of this section can be summarized briefly. The physical type IIB
combination is controlled by Jacobi's identity, while the four sector terms
before the final GSO projection share a common Fourier matrix determined by the
block \eqref{eq:Aseries}. Later sections will act on these sector coefficients
with the Lorentzian and regularized modular-integral prescriptions. At the
present stage, the only purpose is to establish the integrand, the sector
split, and the exact coefficient data on which the later construction is
built.

\section{Cusp regularization}
\label{sec:cusp-regularization}

The Euclidean modular domain supplies the geometric input for the later
regularized mode blocks. A truncated fundamental domain separates into a compact
keyhole region and a cusp strip above height one. On the strip, horizontal
Fourier orthogonality projects a single mode \(q^m\bar q^n\) onto the diagonal
\(m=n\). The Lorentzian prescription and the generalized-exponential-integral
tail both act on this diagonal strip contribution; neither is needed to derive
the Euclidean projection itself.

\subsection{Upper half-plane and fundamental domain}
\label{subsec:cusp-upper-half-plane}

We work on the upper half-plane
\begin{equation}
\HH=\{x+\ii y\in\CC:\; y>0\},
\label{eq:cusp-H}
\end{equation}
with the torus modulus written as
\begin{equation}
\tau=x+\ii y,
\qquad
q=\ee^{2\pi \ii \tau},
\qquad
\bar q=\ee^{-2\pi \ii \bar\tau}.
\label{eq:cusp-tauq}
\end{equation}
The sign in the anti-holomorphic nome is chosen so that the factor \(q^m\bar q^n\) decays exponentially as \(y\to\infty\) whenever \(m+n>0\).

The standard modular fundamental domain for the oriented torus is
\begin{equation}
\cF=
\left\{
\tau\in\HH:
-\frac12\le \operatorname{Re}\tau\le \frac12,
\quad
|\tau|\ge 1
\right\},
\label{eq:cusp-Fcircle}
\end{equation}
which is the usual representative of \(\HH/SL(2,\ZZ)\) in genus-one perturbation theory and in the classical theory of modular forms~\cite{GreenSchwarzWitten1988,Polchinski1998Vol2,Serre1973Course,Apostol1990ModularFunctions,Zagier2008Elliptic}. Since \(y>0\), the condition \(|\tau|\ge 1\) is equivalent inside the vertical strip to
\begin{equation}
y\ge \sqrt{1-x^2}.
\label{eq:cusp-keyholeboundary}
\end{equation}
Thus the same domain may be written as
\begin{equation}
\cF=
\left\{
x+\ii y:
-\frac12\le x\le \frac12,
\quad
y\ge \sqrt{1-x^2}
\right\}.
\label{eq:cusp-Fkeyhole}
\end{equation}
This form displays the lower boundary and isolates the non-compact cusp at
\(y\to\infty\).

\subsection{Truncation at finite height}
\label{subsec:cusp-truncated-domain}

To separate the cusp from the compact part of the domain, we introduce a height cutoff \(Y>1\) and define
\begin{equation}
\cF_Y=\cF\cap\{y\le Y\}.
\label{eq:cusp-FY}
\end{equation}
The restriction \(Y>1\) ensures that the cutoff lies above the unit-circle boundary of the fundamental domain across the full interval \(x\in[-1/2,1/2]\). The truncated domain \(\cF_Y\) is compact and provides the natural Euclidean starting point for later regularization.

For a single Fourier mode we define the truncated mode integral
\begin{equation}
L_{m,n,s}(Y)
=
\int_{\cF_Y}\dd\tau\wedge \dd\bar\tau\;
y^{-s}q^m\bar q^n.
\label{eq:cusp-LmnY}
\end{equation}
At this stage \(m\), \(n\), and \(s\) are fixed parameters, and no subtraction
term is included. In the vacuum application one sets \(s=6\), but the geometric
projection does not require that specialization. The object
\eqref{eq:cusp-LmnY} is only a cutoff Euclidean integral over the truncated
domain.

\subsection{Compact keyhole region}
\label{subsec:cusp-keyhole}

The compact part of the domain is the keyhole region below height one,
\begin{equation}
\cF_1=
\left\{
x+\ii y:
-\frac12\le x\le \frac12,
\quad
\sqrt{1-x^2}\le y\le 1
\right\}.
\label{eq:cusp-F1}
\end{equation}
Its associated mode integral is
\begin{equation}
K_{m,n,s}
=
\int_{\cF_1}\dd\tau\wedge \dd\bar\tau\;
y^{-s}q^m\bar q^n.
\label{eq:cusp-Kmn}
\end{equation}
Because both coordinates are bounded on \(\cF_1\), this integral is finite for fixed \((m,n,s)\). The compact keyhole region is therefore not the source of any cusp sensitivity; that role belongs entirely to the strip above height one.

\subsection{Cusp strip and domain decomposition}
\label{subsec:cusp-strip-decomposition}

The complementary strip between heights one and \(Y\) is
\begin{equation}
S_{1,Y}=
\left\{
x+\ii y:
-\frac12\le x\le \frac12,
\quad
1\le y\le Y
\right\}.
\label{eq:cusp-S1Y}
\end{equation}
Up to the common boundary at \(y=1\), which has measure zero for the two-dimensional integral, the truncated domain decomposes as
\begin{equation}
\cF_Y=\cF_1\cup S_{1,Y}.
\label{eq:cusp-FYsplit}
\end{equation}
This follows immediately from \eqref{eq:cusp-Fkeyhole}: for each fixed \(x\) in the strip, the allowed range in \(\cF_Y\) is \(\sqrt{1-x^2}\le y\le Y\), which is the union of the compact interval \(\sqrt{1-x^2}\le y\le 1\) and the strip interval \(1\le y\le Y\).

We define the strip contribution of a single mode by
\begin{equation}
R_{m,n,s}(Y)
=
\int_{S_{1,Y}}\dd\tau\wedge \dd\bar\tau\;
y^{-s}q^m\bar q^n.
\label{eq:cusp-RmnY}
\end{equation}
The decomposition \eqref{eq:cusp-FYsplit} then implies
\begin{equation}
L_{m,n,s}(Y)=K_{m,n,s}+R_{m,n,s}(Y).
\label{eq:cusp-LKR}
\end{equation}
At this stage the decomposition is purely Euclidean, and all dependence on the
cutoff height \(Y\) is localized in the strip term.

\subsection{Wedge measure and Fourier projection}
\label{subsec:cusp-projection}

The wedge measure convention used throughout the calculation is
\begin{equation}
\dd\tau\wedge \dd\bar\tau=-2\ii\,\dd x\,\dd y.
\label{eq:cusp-wedge}
\end{equation}
Substituting \eqref{eq:cusp-tauq} into a Fourier mode gives
\begin{equation}
q^m\bar q^n
=
\ee^{2\pi \ii (m-n)x}\,
\ee^{-2\pi (m+n)y}.
\label{eq:cusp-modefactor}
\end{equation}
The \(x\)- and \(y\)-dependences therefore factorize on the strip, which is the basic reason the cusp contribution is easy to analyze before any regularization is applied.

Assume from now on that \(m-n\in\ZZ\), as happens in the Fourier expansion of the torus integrand. Then the horizontal integral is the standard Fourier projection on the unit interval,
\begin{equation}
\int_{-1/2}^{1/2}\dd x\,
\ee^{2\pi \ii (m-n)x}
=
\delta_{m,n}.
\label{eq:cusp-xprojection}
\end{equation}
If \(m=n\), the integral is one. If \(m-n=k\in\ZZ\setminus\{0\}\), then
\begin{equation}
\int_{-1/2}^{1/2}\dd x\,\ee^{2\pi \ii kx}
=
\frac{\ee^{\pi \ii k}-\ee^{-\pi \ii k}}{2\pi \ii k}
=
\frac{\sin(\pi k)}{\pi k}
=
0.
\label{eq:cusp-integer-projection}
\end{equation}
Equation~\eqref{eq:cusp-xprojection} is therefore exact and shows that the strip contribution can only survive on diagonal modes.

Substituting \eqref{eq:cusp-wedge} and \eqref{eq:cusp-modefactor} into \eqref{eq:cusp-RmnY} gives
\begin{equation}
R_{m,n,s}(Y)
=
-2\ii
\int_{-1/2}^{1/2}\dd x
\int_1^Y\dd y\;
y^{-s}\,
\ee^{2\pi \ii (m-n)x}
\ee^{-2\pi (m+n)y}.
\label{eq:cusp-Rbeforeproj}
\end{equation}
Using the projection \eqref{eq:cusp-xprojection}, only the diagonal part remains. On the diagonal \(m=n\), the exponential becomes \(\ee^{-4\pi my}\), and the strip contribution reduces to
\begin{equation}
R_{m,n,s}(Y)
=
-2\ii\,\delta_{m,n}
\int_1^Y\dd y\;
y^{-s}\ee^{-4\pi my}.
\label{eq:cusp-Rmain}
\end{equation}
This is the main result of the section. It shows that every off-diagonal mode has vanishing strip contribution, while every diagonal mode is controlled by a one-dimensional tail integral over the cusp direction.

Equation \eqref{eq:cusp-Rmain} is the endpoint of the Euclidean analysis. No
regularized mode block, generalized exponential integral, or Lorentzian cycle
has entered. The result is the geometric input needed later: the cutoff domain
splits into a compact keyhole part and a cusp strip, and the strip acts
diagonally on the Fourier labels.

\section{Lorentzian prescription}
\label{sec:lorentzian-prescription}

The Euclidean cusp strip gives a diagonal long-tube contribution, but the
Euclidean integral alone is not the physical one-loop prescription. The
non-separating cusp corresponds to propagation over a long proper time, and the
worldsheet integral must be interpreted with the string-theoretic analogue of
the Feynman \(i\varepsilon\) prescription. The required replacement is a
Lorentzian continuation of the long-tube variable. Its action on a single
diagonal Fourier mode is the contour-level input from which the regularized
mode blocks will be evaluated.

\subsection{Euclidean strip and Lorentzian continuation}
\label{subsec:lor-cusp-continuation}

In the strip region \(y\ge 1\), the torus develops a long cylinder whose length is measured by the proper-time variable \(y=\operatorname{Im}\tau\). This is the worldsheet regime in which intermediate string states propagate for a long time. Witten's formulation of the string-theoretic \(i\varepsilon\) prescription emphasizes that the physical Lorentzian amplitude is not obtained by keeping this proper time on the positive real axis all the way to infinity~\cite{Witten2013IEpsilon}. Instead, after a sufficiently large Euclidean segment, the contour must be continued into a complex direction that implements the same causal choice as the field-theory Feynman prescription.

The same degeneration region is where modern worldsheet-cut constructions
detect the discontinuity structure of one-loop amplitudes
~\cite{EberhardtMizera2022Cuts}. From the present vacuum point of view, this
means that the cusp strip cannot be treated as an arbitrary Euclidean tail if
the regularized modular integral is to remain compatible with the Lorentzian
definition of string amplitudes. The Euclidean domain decomposition identifies
the strip that must be replaced by a physically meaningful continuation.

The vacuum problem uses only a narrow specialization of the general contour
story. Witten's prescription is formulated for the full one-loop integration
cycle in the presence of general degeneration loci~\cite{Witten2013IEpsilon}.
Here there are no puncture moduli and no Koba--Nielsen kernel. The only modulus
that needs a causal prescription is the long-tube parameter itself. The
calculation therefore reduces to the cusp-level contour object that survives in
the vacuum case.

\subsection{The long-tube continuation}
\label{subsec:lor-long-tube}

Let \(T_0>1\) be a large reference height in the cusp. The Euclidean proper-time segment from \(1\) to \(T_0\) is retained on the real axis, but beyond \(T_0\) the contour is continued vertically into the complex plane. Concretely, we introduce the cusp contour
\begin{equation}
\Gamma(T_0)
=
[1,T_0]\cup\{\,T_0+\ii t:\; t\ge 0\,\}.
\label{eq:lor-GammaT0}
\end{equation}
The first piece is the Euclidean segment, while the second piece is the Lorentzian tail. In the variable \(y\) this is the simplest contour that matches the string-theoretic \(i\varepsilon\) prescription at the non-separating cusp: one proceeds along real Euclidean proper time up to a large cutoff and then turns upward into the imaginary direction~\cite{Witten2013IEpsilon,ManschotWang2024,WangThesis2025}.

The parameter \(T_0\) is not an additional regularization parameter. It marks
the height at which the contour is decomposed into a compact Euclidean segment
and a Lorentzian tail. Different choices of \(T_0\) give different
decompositions of the same contour prescription. The final regularized block is
therefore independent of \(T_0\), although the intermediate Euclidean and
Lorentzian contributions separately depend on it.
Appendix~\ref{app:contour-deformation} gives the corresponding
contour-independence derivation.

At the level of the degeneration coordinate
\begin{equation}
q=\ee^{2\pi \ii \tau},
\label{eq:lor-qdef}
\end{equation}
the Euclidean strip approaches the degeneration point \(q=0\) radially, whereas the Lorentzian continuation turns into a contour of fixed modulus and varying phase. On the vertical segment \(y=T_0+\ii t\),
\begin{equation}
q=\ee^{2\pi\ii x}\ee^{-2\pi T_0}\ee^{-2\pi\ii t}.
\label{eq:lor-q-cycle}
\end{equation}
Thus the contour \eqref{eq:lor-GammaT0} is the proper-time representative of a
cycle around the degeneration point in the complexified \(q\)-plane. This is
the cusp-level realization of the physical \(i\varepsilon\) choice.

\subsection{Single-mode contour decomposition}
\label{subsec:lor-single-mode}

The Euclidean strip acts only on diagonal modes. For a diagonal term \(m=n\),
the strip integrand reduces to
\begin{equation}
y^{-s}\ee^{-4\pi my}.
\label{eq:lor-diagmode}
\end{equation}
The Euclidean tail and the prescribed Lorentzian contour are distinct objects.
The ordinary Euclidean tail is
\begin{equation}
T^E_{m,s}
=
\int_1^\infty \dd y\, y^{-s}\ee^{-4\pi my}.
\label{eq:lor-euclidean-tail}
\end{equation}
The corresponding prescribed strip contour is
\begin{equation}
T^{\Gamma}_{m,s}(T_0)
=
\int_{\Gamma(T_0)}\dd z\, z^{-s}\ee^{-4\pi mz},
\label{eq:lor-lorentzian-tail}
\end{equation}
where \(z\) is the complexified proper-time variable.  Equation
\eqref{eq:lor-lorentzian-tail} is a definition of the contour contribution
associated with the prescription before it is split into its finite Euclidean
part and its vertical tail.

Writing the second segment of \eqref{eq:lor-GammaT0} as \(z=T_0+\ii t\) with \(t\ge 0\), one obtains
\begin{equation}
\int_{\Gamma(T_0)}\dd z\, z^{-s}\ee^{-4\pi mz}
=
\int_1^{T_0}\dd y\, y^{-s}\ee^{-4\pi my}
\;+\;
\ii\int_0^\infty \dd t\,
(T_0+\ii t)^{-s}\ee^{-4\pi m(T_0+\ii t)}.
\label{eq:lor-GammaSplit}
\end{equation}
The first term is the finite Euclidean segment, and the second term is the
Lorentzian vertical contribution.  The vertical tail itself is
\begin{equation}
T^{\rm vert}_{m,s}(T_0)
=
\int_{T_0}^{T_0+\ii\infty}\dd z\,z^{-s}\ee^{-4\pi mz}.
\label{eq:lor-vertical-tail}
\end{equation}
This decomposition has two structural
consequences. First, the prescription is diagonal mode by mode because the
horizontal strip projection has already produced \(\delta_{m,n}\). Second, the
Lorentzian contribution is complex even when the Euclidean integrand is real;
the contour therefore carries the analytic data that later appears as the
imaginary part of the regularized modular integral.

The formula \eqref{eq:lor-GammaSplit} is the vacuum specialization of the
general contour prescription. In a non-vacuum amplitude the same continuation
coexists with puncture moduli, worldsheet Green functions, and kinematic branch
structure. In the vacuum problem these data are absent, and the only remaining
analytic issue is the cusp tail of a diagonal Fourier mode.

\subsection{From Euclidean strip to Lorentzian strip contribution}
\label{subsec:lor-stripreplacement}

Combining the diagonal strip projection with the vertical tail
\eqref{eq:lor-vertical-tail}, the Lorentzian tail contribution that enters the
regularized block is
\begin{equation}
R^{\rm Lor}_{m,n,s}(T_0)
=
-2\ii\,\delta_{m,n}
\int_{T_0}^{T_0+\ii\infty}\dd z\, z^{-s}\ee^{-4\pi mz}.
\label{eq:lor-Rlor}
\end{equation}
Equivalently,
\begin{equation}
R^{\rm Lor}_{m,n,s}(T_0)
=
-2\ii\,\delta_{m,n}
\left[
\ii\int_0^\infty \dd t\,
(T_0+\ii t)^{-s}\ee^{-4\pi m(T_0+\ii t)}
\right].
\label{eq:lor-RlorSplit}
\end{equation}
This contour expression is the Lorentzian-prescribed counterpart of the
diagonal cusp tail after the finite Euclidean segment has been included in the
compact keyhole domain.  Evaluating \eqref{eq:lor-Rlor} in a form adapted to
Fourier-mode regularization gives the generalized exponential integral and
turns the contour-level object into a special-function block that can be summed
against the auxiliary-sector Fourier coefficients. The construction separates
the Euclidean projection, the vertical Lorentzian contour, and the resulting
special-function evaluation.

\section{Regularized mode integrals}
\label{sec:regularized-mode-integrals}

The cusp contribution has two distinct pieces: a compact keyhole integral that
is present for every Fourier mode, and a long-tube contribution that survives
only after horizontal projection onto the diagonal \(m=n\). The Lorentzian
contour fixes the latter contribution, while the generalized exponential
integral gives its explicit mode-by-mode value. This produces the regularized
block \(L^r_{m,n,s}\), the object that will be paired with the type IIB sector
Fourier coefficients.

The full contour equivalence between the string-theoretic \(i\varepsilon\)
prescription and the \(E_s\)-regularized modular-integral representation was
developed by Manschot and Wang~\cite{ManschotWang2024} and in Wang's detailed
treatment~\cite{WangThesis2025}. The vacuum calculation uses
the specialization in which the relevant diagonal arguments satisfy
\(4\pi m\ge0\) and the cutoff is fixed to the canonical value \(T_0=1\).

\subsection{Generalized exponential integrals}
\label{subsec:reg-generalized-exponential-integrals}

The generalized exponential integral is normalized as
\begin{equation}
E_s(z)=\int_1^\infty \ee^{-zt}t^{-s}\dd t,
\qquad
\Re(s)>1.
\label{eq:reg-Esdef}
\end{equation}
For \(z\) on the positive real axis this is equivalent to the incomplete-gamma
representation employed in the modular-regularization literature,
\begin{equation}
E_s(z)=z^{\,s-1}\int_z^\infty \ee^{-u}u^{-s}\dd u,
\label{eq:reg-Esgamma}
\end{equation}
obtained from the substitution \(u=zt\)~\cite{ManschotWang2024,WangThesis2025,Iwaniec2002Spectral}.
The first form is the most convenient one for the vacuum discussion, because
the strip tail is already written as an integral over \(y\ge1\); the second
form makes the relation to the standard special-function literature explicit.

The constant mode is included by continuity:
\begin{equation}
E_s(0)=\frac{1}{s-1},
\qquad
\Re(s)>1.
\label{eq:reg-Eszero}
\end{equation}
In particular, for the type IIB torus vacuum where \(s=6\),
\begin{equation}
E_6(0)=\frac15.
\label{eq:reg-E60}
\end{equation}
This value will later control the constant-mode contribution, but for the
moment it is simply the \(m=0\) member of the same special-function family that
captures all diagonal cusp tails.

\subsection{From the Lorentzian contour to the \texorpdfstring{$E_s$}{Es} tail}
\label{subsec:reg-cusp-tail}

After the compact keyhole region has been cut off at height \(T_0\), the
remaining cusp contribution is the vertical part of the Lorentzian contour,
\begin{equation}
\Gamma_\infty(T_0)=\{\,T_0+\ii t:\;t\ge0\,\}.
\label{eq:reg-vertical-tail}
\end{equation}
The contour-level tail entering the regularized block is therefore
\begin{equation}
R^{\rm Lor}_{m,n,s}(T_0)
=
-2\ii\,\delta_{m,n}
\int_{T_0}^{T_0+\ii\infty}\dd z\, z^{-s}\ee^{-4\pi mz}.
\label{eq:reg-Rlor}
\end{equation}
This differs from the full contour \(\Gamma(T_0)\) of
Section~\ref{sec:lorentzian-prescription}, which also contains the finite
Euclidean segment from \(1\) to \(T_0\).  That finite segment is already part of
the compact domain \(\cF_{T_0}\).  The \(E_s\) term in the block contains only
the remaining vertical tail.

For the non-negative diagonal modes that occur in the Type IIB vacuum,
Appendix~\ref{app:contour-deformation} proves
\begin{equation}
\int_{T_0}^{T_0+\ii\infty}\dd z\,z^{-s}\ee^{-4\pi mz}
=
T_0^{\,1-s}E_s(4\pi mT_0),
\qquad
m\ge0,\quad \Re(s)>1.
\label{eq:reg-tail-lemma}
\end{equation}
This is the vacuum specialization of the Lorentzian/\(E_s\) relation used in
the regularized modular-integral framework of
Manschot and Wang~\cite{ManschotWang2024,WangThesis2025}.  The same local
contour component is the standard-cusp ingredient that enters
Rademacher-type reorganizations of modular integrals
~\cite{Rademacher1938Partition,Niebur1973Nonanalytic,BacciantiChandraEberhardtHartmanMizera2025}.
For the present sector functional it is represented by the vertical tail in
\eqref{eq:reg-tail-lemma}.
Combining \eqref{eq:reg-Rlor} with \eqref{eq:reg-tail-lemma} gives the analytic
special-function tail
\begin{equation}
-2\ii\,\delta_{m,n}\,T_0^{\,1-s}E_s(4\pi mT_0).
\label{eq:reg-T0tail}
\end{equation}
The parameter \(T_0\) is the contour splitting height introduced earlier. It
is not a new regularization scheme parameter. Different values of \(T_0\)
simply correspond to different ways of decomposing the same contour into a
compact Euclidean part and a Lorentzian tail;
Appendix~\ref{app:contour-deformation} derives the corresponding
cutoff-independence relation.

The canonical vacuum choice is
\begin{equation}
T_0=1.
\label{eq:reg-T0one}
\end{equation}
At this value, \eqref{eq:reg-T0tail} reduces to
\begin{equation}
-2\ii\,\delta_{m,n}E_s(4\pi m).
\label{eq:reg-unit-tail}
\end{equation}
This is exactly the analytic remnant of the cusp strip after the compact
keyhole region has been separated off. The factor \(\delta_{m,n}\) is inherited
from the horizontal strip projection,
while the overall factor \(-2\ii\) comes from the wedge-measure convention
\(\dd\tau\wedge\dd\bar\tau=-2\ii\,\dd x\,\dd y\).

The analytic tail also admits a direct Euclidean finite-strip interpretation.
For \(a=4\pi m\) one has
\begin{equation}
\int_1^\infty \dd y\, y^{-s}\ee^{-ay}=E_s(a),
\label{eq:reg-full-tail}
\end{equation}
and, after the change of variables \(y=Yu\),
\begin{equation}
\int_Y^\infty \dd y\, y^{-s}\ee^{-ay}
=
Y^{1-s}E_s(aY).
\label{eq:reg-cutoff-tail}
\end{equation}
Therefore the Euclidean strip integral between \(1\) and \(Y\) is
\begin{equation}
\int_1^Y \dd y\, y^{-s}\ee^{-4\pi my}
=
E_s(4\pi m)-Y^{1-s}E_s(4\pi mY).
\label{eq:reg-strip-Es}
\end{equation}
This identity displays the three terms that enter the construction: the finite
strip, the special-function tail, and the cutoff remainder.

\subsection{Definition of the regularized mode block}
\label{subsec:reg-Lrdef}

Let \(\cF_1\) be the compact keyhole region of the truncated fundamental
domain. The regularized mode block is
\begin{equation}
L^r_{m,n,s}
=
\int_{\cF_1}\dd\tau\wedge\dd\bar\tau\,
y^{-s}q^m\bar q^n
-2\ii\,\delta_{m,n}E_s(4\pi m).
\label{eq:reg-Lr}
\end{equation}
This is the final block-level object used in the present paper. The first
term is the compact-domain integral, present for every mode. The second term is
the analytic cusp contribution, present only on the diagonal. The subtraction
language of regularization is therefore a shorthand for the underlying
geometric decomposition: the full block is a compact contribution plus a
Lorentzian-prescribed analytic tail.

For the finite-mode evaluation, \eqref{eq:reg-Lr} is written in terms of the
real compact integral
\begin{equation}
I_{\cF_1}^{\rm real}(m,n,s)
=
\int_{-1/2}^{1/2}\dd x
\int_{\sqrt{1-x^2}}^{1}\dd y\;
y^{-s}\ee^{2\pi \ii(m-n)x}\ee^{-2\pi(m+n)y}.
\label{eq:reg-IF1real}
\end{equation}
Since
\begin{equation}
\int_{\cF_1}\dd\tau\wedge\dd\bar\tau\,
y^{-s}q^m\bar q^n
=
-2\ii\,I_{\cF_1}^{\rm real}(m,n,s),
\label{eq:reg-wedgecompact}
\end{equation}
the block can equivalently be written as
\begin{equation}
L^r_{m,n,s}
=
-2\ii\Bigl[
I_{\cF_1}^{\rm real}(m,n,s)
\,+\,
\delta_{m,n}E_s(4\pi m)
\Bigr].
\label{eq:reg-Lrrealform}
\end{equation}
This is the form used in the numerical evaluation.

\subsection{Vacuum specialization and branch convention}
\label{subsec:reg-branch}

For the type IIB torus vacuum one specializes to
\begin{equation}
s=6,
\label{eq:reg-s6again}
\end{equation}
so that
\begin{equation}
L^r_{m,n,6}
=
\int_{\cF_1}\dd\tau\wedge\dd\bar\tau\,
y^{-6}q^m\bar q^n
-2\ii\,\delta_{m,n}E_6(4\pi m).
\label{eq:reg-Lr-six}
\end{equation}
In this vacuum setting the diagonal argument \(4\pi m\) lies on the
non-negative real axis, so the block does not cross the negative-real branch
cut relevant to genuine threshold problems. Nevertheless, the analytic
continuation convention should be fixed already here, because the same
special-function block is the object that later generalizes to polar and
non-vacuum contexts. The principal branch convention is
\begin{equation}
\operatorname{Arg}z\in(-\pi,\pi].
\label{eq:reg-branch}
\end{equation}
With this convention, for \(x>0\),
\begin{equation}
E_N(-x+\ii0)-E_N(-x-\ii0)
=
-\,2\pi\ii\,\frac{x^{N-1}}{\Gamma(N)},
\qquad N\in\ZZ_{\ge1}.
\label{eq:reg-branch-disc}
\end{equation}
Here the discontinuity is upper boundary value minus lower boundary value and
\(\operatorname{Log}(-x+\ii0)=\log x+\ii\pi\).  With the opposite
discontinuity convention, the sign changes.  The vacuum modes considered here
never require \eqref{eq:reg-branch-disc}; the formula fixes the convention of
the special-function block for later non-vacuum applications.

The regularized construction is linear:
\begin{equation}
I^r[f+g]=I^r[f]+I^r[g],
\qquad
I^r[f]=\sum_{m,n}F(m,n)L^r_{m,n,s}.
\label{eq:reg-linearity}
\end{equation}
This linearity allows the GSO projection to be applied after the auxiliary
sector data have been regularized.  Once the diagonal cusp contribution has
been incorporated into \(L^r_{m,n,6}\), the remaining vacuum calculation is
governed by the type IIB sector Fourier coefficients and by the compact-keyhole
integrals appearing in the first term of \eqref{eq:reg-Lr-six}.

\section{Sector amplitudes}
\label{sec:sector-amplitudes}

The sector calculation is the central construction of the paper.  It combines
the Fourier coefficient matrices \(F_{XY}(m,n)\) of the four auxiliary Type IIB
sector terms with the regularized mode block \(L^r_{m,n,6}\). Their pairing
defines an auxiliary sector integral for each \(X,Y\in\{V,S\}\). The signed GSO
contraction is not applied at this stage, so the quantities below are not
physical vacuum amplitudes; they are the sector-level objects on which the
Lorentzian and \(E_s\)-regularized prescriptions are compared.

\subsection{Definition of the sector sums}
\label{subsec:sector-sums}

For each auxiliary sector,
\begin{equation}
\cZ_{XY}(\tau,\bar\tau)
=
\sum_{m,n\ge0}F_{XY}(m,n)q^m\bar q^n.
\label{eq:sector-ZXY-fourier}
\end{equation}
The corresponding regularized sector integral is defined mode by mode by
\begin{equation}
I^r_{XY}
=
\sum_{m,n\ge0}F_{XY}(m,n)L^r_{m,n,6}.
\label{eq:sector-Ir}
\end{equation}
This sector integral is the primary regularized quantity computed in the
vacuum problem. The index \(6\) is not an adjustable regulator; it is fixed by
the measure \(d\tau\wedge d\bar\tau\,y^{-6}\) in the original torus vacuum
integral \eqref{eq:IIBvacuum}.

The auxiliary numerical value uses the amplitude convention
\begin{equation}
\cA_{XY}=\ii I^r_{XY}.
\label{eq:sector-Adef}
\end{equation}
With the wedge convention \(d\tau\wedge d\bar\tau=-2\ii\,dx\,dy\), the blocks
\(L^r_{m,n,6}\) that occur in the present vacuum problem are naturally
imaginary, and \eqref{eq:sector-Adef} converts the sector value into the real
normalization used for the numerical sector datum.

To separate exact coefficient algebra from numerical convergence, introduce the finite Fourier cutoff
\begin{equation}
I^{r,(M)}_{XY}
=
\sum_{m,n=0}^{M}F_{XY}(m,n)L^r_{m,n,6},
\qquad
\cA^{(M)}_{XY}=\ii I^{r,(M)}_{XY}.
\label{eq:sector-finite-cutoff}
\end{equation}
At finite \(M\), all sector identities below are identities between finite sums. The later limiting process only determines the common value of those sums.

The finite evaluations quoted below use deterministic quadrature for the
compact keyhole integrals, exact generation of the \(SO(8)\) Fourier
coefficients, and the special-function representation of the cusp and compact
pieces through \(E_s\) and incomplete-gamma functions.  The sector sums are
then evaluated with a finite Fourier cutoff, while the same holomorphic
degeneracies may also be reconstructed from the \(\Gamma(2)\) Rademacher
coefficient construction described in Section~\ref{subsec:fourierdata}.

\subsection{Mode expansion into keyhole and cusp terms}
\label{subsec:keyhole-cusp-split}

Substituting the block \eqref{eq:reg-Lr-six} into \eqref{eq:sector-Ir} gives the central sector formula
\begin{equation}
I^r_{XY}
=
\sum_{m,n\ge0}
F_{XY}(m,n)
\int_{\cF_1}
\dd\tau\wedge\dd\bar\tau\,
y^{-6}q^m\bar q^n
-
2\ii
\sum_{m\ge0}
F_{XY}(m,m)E_6(4\pi m).
\label{eq:sector-expanded}
\end{equation}
Define the compact keyhole mode integral
\begin{equation}
K_{m,n}
=
\int_{\cF_1}
\dd\tau\wedge\dd\bar\tau\,
y^{-6}q^m\bar q^n.
\label{eq:sector-K}
\end{equation}
In real variables,
\begin{equation}
K_{m,n}
=
-2\ii
\int_{-1/2}^{1/2}\dd x
\int_{\sqrt{1-x^2}}^1\dd y\,
y^{-6}
\ee^{2\pi\ii(m-n)x}
\ee^{-2\pi(m+n)y}.
\label{eq:sector-K-real}
\end{equation}
The cusp contribution is diagonal:
\begin{equation}
C_m=-2\ii E_6(4\pi m),
\qquad
I^{\rm cusp}_{XY}
=
\sum_{m\ge0}F_{XY}(m,m)C_m.
\label{eq:sector-cusp}
\end{equation}
The diagonal restriction is the same horizontal projection already obtained on
the cusp strip: the \(x\)-integral kills every mode with \(m\ne n\).

\subsection{Diagonal and off-diagonal parts}
\label{subsec:diag-off}

The full sector sum decomposes into a diagonal part and an off-diagonal part,
\begin{equation}
I^r_{XY}=I^{r,\diag}_{XY}+I^{r,\off}_{XY}.
\label{eq:sector-diag-off-sum}
\end{equation}
The diagonal contribution contains both the compact keyhole integral and the cusp subtraction,
\begin{equation}
I^{r,\diag}_{XY}
=
\sum_{m\ge0}
F_{XY}(m,m)
\left[
K_{m,m}-2\ii E_6(4\pi m)
\right],
\label{eq:sector-diag}
\end{equation}
whereas the off-diagonal contribution is purely compact,
\begin{equation}
I^{r,\off}_{XY}
=
\sum_{\substack{m,n\ge0\\m\ne n}}
F_{XY}(m,n)K_{m,n}.
\label{eq:sector-off}
\end{equation}
The off-diagonal modes have no cusp subtraction because the cusp strip projection kills them.

\subsection{Coefficient-level sector equality}
\label{subsec:sector-equality}

The equality of the four sector amplitudes is not a numerical observation, nor
is it inferred from the final signed projection. It follows directly from the Type IIB
closed-sector character identity \(A_V=A_S\). Write the common block as
\begin{equation}
A_V(\tau)=A_S(\tau)=A(\tau)=\sum_{m\ge0}d_mq^m.
\label{eq:sector-common-block}
\end{equation}
Then every sector coefficient matrix has the same rank-one form,
\begin{equation}
F_{VV}(m,n)=F_{VS}(m,n)=F_{SV}(m,n)=F_{SS}(m,n)=d_md_n.
\label{eq:sector-F-equality}
\end{equation}
The coefficients \(d_m\) may be read directly from the product expansion of
\(A\), or equivalently from the \(\Gamma(2)\) Rademacher reconstruction of the
holomorphic block described in Section~\ref{subsec:fourierdata}
~\cite{ChengDuncan2012,BacciantiChandraEberhardtHartmanMizera2025}.  The sector
functional depends only on the resulting integer degeneracies and on the
regularized mode block \(L^r_{m,n,6}\).  Thus the Rademacher construction
supplies an independent cusp-theoretic description of the coefficient input,
while the Lorentzian/\(E_6\) block remains the analytic pairing that defines
the sector integral.
This identity holds at the level of Fourier data before integration.
Substituting \eqref{eq:sector-F-equality} into the finite sum
\eqref{eq:sector-finite-cutoff} gives
\begin{equation}
I^{r,(M)}_{VV}
=
I^{r,(M)}_{VS}
=
I^{r,(M)}_{SV}
=
I^{r,(M)}_{SS}
\qquad
\text{for every }M\ge0,
\label{eq:sector-finite-equality}
\end{equation}
and hence also
\begin{equation}
\cA^{(M)}_{VV}
=
\cA^{(M)}_{VS}
=
\cA^{(M)}_{SV}
=
\cA^{(M)}_{SS}.
\label{eq:sector-finite-A-equality}
\end{equation}

Passing to the all-mode regularized sums gives
\begin{equation}
I^r_{VV}=I^r_{VS}=I^r_{SV}=I^r_{SS}.
\label{eq:sector-I-equality}
\end{equation}
The same equality holds for the amplitudes \(\cA_{XY}\). The proof is
coefficient-level and term-by-term; Appendix~\ref{app:numerical-details}
gives the finite all-mode value of this shared sector integral. The physical
signed contraction
\[
I^r_{VV}-I^r_{VS}-I^r_{SV}+I^r_{SS}
\]
is a later projection of the sector result.

\section{Constant mode}
\label{sec:constant-mode}

The constant Fourier mode can be evaluated analytically before the full sector
sum is considered. It gives the cleanest exact normalization datum in the
vacuum problem: unlike the full sector value, the zero-mode contribution
follows directly from the compact keyhole geometry and the constant cusp tail in
the regularized block~\cite{ManschotWang2024,WangThesis2025}. It also fixes the
normalization distinction between the closed-string coefficient \(64\) and the
holomorphic comparison coefficient \(16\).

\subsection{The constant mode block}
\label{subsec:constant-block}

For the constant mode,
\begin{equation}
m=n=0,
\qquad
q^0\bar q^0=1.
\label{eq:mneqzero}
\end{equation}
The general block \eqref{eq:reg-Lr-six} becomes
\begin{equation}
L^r_{0,0,6}
=
\int_{\cF_1}\dd\tau\wedge\dd\bar\tau\,y^{-6}
-2\ii E_6(0).
\label{eq:L0006}
\end{equation}
Using the real compact integral
\begin{equation}
I_{\cF_1}^{\rm real}(0,0,6)
=
\int_{-1/2}^{1/2}\dd x
\int_{\sqrt{1-x^2}}^{1}\dd y\,y^{-6},
\label{eq:IF1real00}
\end{equation}
and the wedge-measure convention
\begin{equation}
\dd\tau\wedge\dd\bar\tau=-2\ii\,\dd x\,\dd y,
\label{eq:wedgeagain}
\end{equation}
one obtains
\begin{equation}
L^r_{0,0,6}
=
-2\ii\Bigl[
I_{\cF_1}^{\rm real}(0,0,6)+E_6(0)
\Bigr].
\label{eq:L0006real}
\end{equation}
Thus the zero-mode problem reduces to a compact real integral plus the constant cusp tail.

\subsection{Zero-mode finite part}
\label{subsec:keyhole-integral}

The compact contribution is best viewed as the \(m=n=0\) specialization of the
same finite-part functional used for the nonzero modes.  For \(\Re(s)>1\),
the real compact term is
\begin{equation}
\begin{aligned}
I_{\cF_1}^{\rm real}(0,0,s)
\,&=
\frac{1}{s-1}
\left[
\int_{-1/2}^{1/2}
(1-x^2)^{(1-s)/2}\dd x
-1
\right]
\\
&=
\frac{2}{s-1}
\int_0^{\pi/6}\cos^{2-s}\theta\,\dd\theta
-
\frac{1}{s-1}.
\end{aligned}
\label{eq:IF1-general-s}
\end{equation}
The lower boundary \(y=\sqrt{1-x^2}\) is the circular arc of the modular
keyhole, while the upper boundary \(y=1\) is the split between the compact
domain and the cusp strip.  Thus \eqref{eq:IF1-general-s} is not a separate
rule for the constant mode; it is the compact finite part before the diagonal
cusp tail is added.

For the vacuum measure \(s=6\), the remaining elementary integral gives
\begin{equation}
I_{\cF_1}^{\rm real}(0,0,6)
=
\frac15\left(\frac{20}{9\sqrt3}-1\right).
\label{eq:IF1exact}
\end{equation}

\subsection{Cusp tail and cutoff independence}
\label{subsec:cusp-tail}

The \(E_s(0)\) contribution is the constant diagonal cusp tail.  If the cusp
strip is split at height \(Y\ge1\), this tail is
\begin{equation}
T_s(Y)
=
\int_Y^\infty y^{-s}\dd y
=
\frac{Y^{1-s}}{s-1}.
\label{eq:constant-tail-Y}
\end{equation}
Changing \(Y\) transfers a boundary strip between the compact contribution and
the tail; the sum is independent of that intermediate split.  In the
normalization of the regularized block the split is \(Y=1\), and therefore
\begin{equation}
T_s(1)=E_s(0)=\frac{1}{s-1}.
\label{eq:Eszeroagain}
\end{equation}
For \(s=6\), this gives \(E_6(0)=1/5\).  Combining the compact finite part
with the constant cusp tail gives
\begin{equation}
I_{\cF_1}^{\rm real}(0,0,6)+E_6(0)
=
\frac{4}{9\sqrt3}.
\label{eq:compactpluscusp}
\end{equation}
Substitution into \eqref{eq:L0006real} yields the exact regularized zero-mode
block
\begin{equation}
L^r_{0,0,6}
=
-2\ii\left(\frac{4}{9\sqrt3}\right)
=
-\frac{8\ii}{9\sqrt3}.
\label{eq:L0006exact}
\end{equation}
The block is now ready to be paired with the closed-sector Fourier
coefficient.

\subsection{From the block to the sector amplitude}
\label{subsec:block-to-amplitude}

The relevant closed-string coefficient is the left-right product
\begin{equation}
F_{XY}(0,0)=8\times 8=64.
\label{eq:F00again}
\end{equation}
This is the coefficient already fixed in \eqref{eq:F00}. It should not be confused with the holomorphic comparison series
\begin{equation}
\frac{\vartheta_2(\tau)^4}{\eta(\tau)^{12}}
=
16+256q+\cO(q^2),
\label{eq:holomorphic16}
\end{equation}
which belongs to a single holomorphic block and is familiar from open or
unoriented one-loop normalization comparisons~\cite{AngelantonjSagnotti2002,WangThesis2025}. In the closed type IIB sector the block used here is instead
\begin{equation}
A(\tau)=\frac{\vartheta_2(\tau)^4}{2\eta(\tau)^{12}}
=
8+128q+\cO(q^2),
\label{eq:AV8}
\end{equation}
so the sector coefficient is \(8\cdot 8\), not \(16\).

Using the amplitude convention \eqref{eq:sector-Adef}, the constant-mode contribution in one auxiliary sector is
\begin{equation}
\cA^{XY}_{00}
=
\ii\,F_{XY}(0,0)L^r_{0,0,6}.
\label{eq:A00def}
\end{equation}
Equivalently, by \eqref{eq:L0006real},
\begin{equation}
\cA^{XY}_{00}
=
2F_{XY}(0,0)
\Bigl[
I_{\cF_1}^{\rm real}(0,0,6)+E_6(0)
\Bigr].
\label{eq:A00structural}
\end{equation}
Finally, using \eqref{eq:compactpluscusp} and \eqref{eq:F00again},
\begin{equation}
\cA^{XY}_{00}
=
2\cdot64\cdot\frac{4}{9\sqrt3}
=
\frac{512}{9\sqrt3}.
\label{eq:A00final}
\end{equation}
This is an exact single-sector zero-mode normalization datum. It is not the full sector value, because non-constant modes have not been included, and it is not a physical type IIB vacuum energy, because the physical GSO combination is treated separately.

\section{GSO projection of the sector functionals}
\label{sec:gso-projection}

The regularized quantities constructed in Section~\ref{sec:sector-amplitudes}
are sector functionals.  The present section applies the physical Type IIB GSO
projection to those sector functionals.  Its role is not to introduce a new
sector value, but to show that the sector-level regularization is compatible
with the standard supersymmetric cancellation after the signed spin-structure
contraction is imposed.

\subsection{Auxiliary sectors and coefficient identities}
\label{subsec:gso-coefficients}

The sector-resolved integrands \(\cZ_{VV}\), \(\cZ_{VS}\), \(\cZ_{SV}\), and
\(\cZ_{SS}\) enter before the physical spin-structure contraction and are
therefore auxiliary. The physical type IIB torus integrand is the signed
combination
\(\cZ_{\rm IIB}=\cZ_{VV}-\cZ_{VS}-\cZ_{SV}+\cZ_{SS}\), which is the sector
expansion of \((V_8-S_8)(\bar V_8-\bar S_8)/|\eta|^{16}\).  The Jacobi
identity \(V_8=S_8\) is used below in its coefficient-level form, after the
regularized sector sums have been defined. Each sector has the Fourier
expansion
\begin{equation}
\cZ_{XY}(\tau,\bar\tau)
=
\sum_{m,n\ge0}F_{XY}(m,n)q^m\bar q^n.
\label{eq:gso-sector-fourier}
\end{equation}
The holomorphic identity \(A_V=A_S\) gives
\begin{equation}
F_{VV}(m,n)
=
F_{VS}(m,n)
=
F_{SV}(m,n)
=
F_{SS}(m,n)
\label{eq:gso-F-equality}
\end{equation}
for every \(m,n\). Define the GSO-signed coefficient
\begin{equation}
F_{\rm GSO}(m,n)
=
F_{VV}(m,n)-F_{VS}(m,n)-F_{SV}(m,n)+F_{SS}(m,n).
\label{eq:gso-F-def}
\end{equation}
Substitution of \eqref{eq:gso-F-equality} gives
\begin{equation}
F_{\rm GSO}(m,n)=0
\qquad
\text{for all }m,n\ge0.
\label{eq:gso-F-zero}
\end{equation}
This is the coefficient-level form of the supersymmetric cancellation. It precedes any integration, any cutoff, and any numerical approximation.

Equivalently, with signs
\begin{equation}
\sigma_V=+1,
\qquad
\sigma_S=-1,
\label{eq:gso-signs}
\end{equation}
one may write
\begin{equation}
F_{\rm GSO}(m,n)
=
\sum_{X,Y\in\{V,S\}}\sigma_X\sigma_YF_{XY}(m,n)
=
\bigl[a_V(m)-a_S(m)\bigr]\bigl[a_V(n)-a_S(n)\bigr].
\label{eq:gso-factorized}
\end{equation}
The final expression is zero because \(a_V(k)=a_S(k)\) for every \(k\). Thus
the cancellation is an identity in the Fourier data and is independent of any
finite-cutoff sector evaluation.

\subsection{Commutativity with the regularized sum}
\label{subsec:gso-commutativity}

The regularized sector integrals are
\begin{equation}
I^r_{XY}
=
\sum_{m,n\ge0}F_{XY}(m,n)L^r_{m,n,6}.
\label{eq:gso-sector-Ir}
\end{equation}
Because \(L^r_{m,n,6}\) is the same block for all four sectors, the physical signed integral can be formed after the sector regularization:
\begin{equation}
I^r_{\rm IIB}
=
I^r_{VV}-I^r_{VS}-I^r_{SV}+I^r_{SS}
=
\sum_{m,n\ge0}F_{\rm GSO}(m,n)L^r_{m,n,6}.
\label{eq:gso-I-sum}
\end{equation}
Using \eqref{eq:gso-F-zero},
\begin{equation}
I^r_{\rm IIB}=0.
\label{eq:gso-I-zero}
\end{equation}
Equation~\eqref{eq:gso-I-zero} expresses the commutativity between modular
regularization and the type IIB GSO projection in the vacuum problem. The
Fourier coefficients may be contracted before regularization, giving zero mode
by mode, or the auxiliary sectors may be regularized first and contracted
afterwards.

With the amplitude convention \(\cA_{XY}=\ii I^r_{XY}\), the same identity gives
\begin{equation}
\cA_{\rm IIB}
=
\cA_{VV}-\cA_{VS}-\cA_{SV}+\cA_{SS}
=0.
\label{eq:gso-amplitude-zero}
\end{equation}

\subsection{Diagonal projector for cusp tails}
\label{subsec:gso-diagonal-discontinuity}

The cusp part of the regularized block is more specific than the full signed
vacuum functional: only the diagonal tail survives the horizontal strip
projection.
Let \(D\) denote the diagonal projector on Fourier coefficients,
\begin{equation}
(DF)(m)=F(m,m).
\label{eq:gso-diagonal-projector}
\end{equation}
The cusp part of the regularized sector functional can then be written as
\begin{equation}
I^{r,\mathrm{cusp}}_{XY}
=
-2\ii
\sum_{m\ge0}
(DF_{XY})(m)E_6(4\pi m).
\label{eq:gso-sector-cusp-functional}
\end{equation}
The GSO contraction commutes with \(D\), since both operations act only on the
coefficient array:
\begin{equation}
D F_{\rm GSO}
=
D(F_{VV}-F_{VS}-F_{SV}+F_{SS})
=
DF_{VV}-DF_{VS}-DF_{SV}+DF_{SS}.
\label{eq:gso-D-commutes}
\end{equation}
Using \eqref{eq:gso-F-zero}, this gives
\begin{equation}
(D F_{\rm GSO})(m)=F_{\rm GSO}(m,m)=0.
\label{eq:gso-diag-zero}
\end{equation}
Hence the diagonal cusp contribution to the physical signed functional is
\begin{equation}
\begin{aligned}
I^{r,\mathrm{cusp}}_{\rm IIB}
&=
-2\ii\sum_{m\ge0}F_{\rm GSO}(m,m)E_6(4\pi m)
\\
&=0.
\end{aligned}
\label{eq:gso-cusp-zero}
\end{equation}
This coefficient-level identity is the input relevant to the \(E_s\) tail; it
is stronger than a cancellation of final numerical sector values.

No negative-axis branch contribution is used in the proof above.  The Type IIB
vacuum sector has non-polar support \(m,n\ge0\), so the diagonal argument
\(4\pi m\) lies on the non-negative real axis before the GSO contraction is
applied.  If one formally continues the same diagonal block to a negative
argument, the convention in Section~\ref{subsec:reg-branch} gives, for \(x>0\),
\begin{equation}
\operatorname{Disc}E_6(-x)
:=
E_6(-x+\ii0)-E_6(-x-\ii0)
=
-\,2\pi\ii\,\frac{x^5}{\Gamma(6)},
\label{eq:gso-Es-disc}
\end{equation}
where the discontinuity is the upper boundary value minus the lower boundary
value.  For a formally continued diagonal mode with \(m<0\), one has
\(x=-4\pi m\).  The discontinuity of the block would then be
\begin{equation}
\operatorname{Disc}L^r_{m,n,6}
=
-2\ii\,\delta_{m,n}\,\operatorname{Disc}E_6(4\pi m)
=
-4\pi\,\delta_{m,n}\frac{(-4\pi m)^5}{\Gamma(6)}
\qquad (m<0).
\label{eq:gso-Lr-disc}
\end{equation}
For this continued expression, a signed discontinuity would be
\begin{equation}
\operatorname{Disc}I^r_{\rm IIB}
=
\sum_{m<0}F_{\rm GSO}(m,m)\,
\operatorname{Disc}L^r_{m,m,6}.
\label{eq:gso-signed-disc}
\end{equation}
This last formula is not used in the vacuum calculation.  In the actual vacuum
sector treated here the support condition \(m,n\ge0\) makes the formal
negative-mode sum empty before any physical conclusion is drawn.  The only
active input needed in the proof is the diagonal coefficient identity
\eqref{eq:gso-diag-zero}.

\section{Summary and Discussions}
\label{sec:discussion}

We have formulated the Type IIB torus vacuum as a regularized modular
calculation on the sector-resolved Fourier data of the oriented closed torus.
The construction separates three ingredients that are often compressed in the
final supersymmetric answer: the Lorentzian prescription at the non-separating
cusp, its equivalent \(E_s\)-regularized mode block, and the algebraic GSO
contraction of the four spin-structure sectors.  The sector functionals provide
a definite closed-string test case for the regularized long-tube functional
before the signed projection is taken.

Each torus sector is paired with the \(E_s\)-regularized block
\(L^r_{m,n,6}\), using the Type IIB closed left-right coefficients
\(F_{XY}(m,n)\), before the physical GSO projection is applied.  This
construction gives the exact zero-mode normalization
\[
\cA^{XY}_{00}=\frac{512}{9\sqrt3}
\]
and the finite all-mode sector value given in
Appendix~\ref{app:numerical-details}.  The subsequent Type IIB GSO contraction
is the coefficient-level implementation of the Jacobi identity in the
\(SO(8)\)-character organization of the genus-one torus integrand
~\cite{GreenSchwarzWitten1988,Polchinski1998Vol2}.

Appendix~\ref{app:unified-lorentz-es} packages the construction as an identity
of sector functionals.  The non-polar vacuum expansion, uniform convergence,
and zero-mode projector allow the sector Fourier series to be inserted into
the same regularized functional, while the non-negative contour lemma
identifies the Lorentzian vertical tail with the corresponding \(E_s\) tail
after specialization to the vacuum measure.  The sector values are the objects
on which the Lorentzian and \(E_s\)-regularized prescriptions are compared
before the physical projection is applied.  The exact physical projection gives
\[
I^r_{\rm IIB}=0,
\qquad
\cA_{\rm IIB}=0.
\]
The sector values are regularized data for the four spin-structure components
of the oriented Type IIB torus integrand.  Since the vacuum has no puncture
integrations and no Koba--Nielsen factor, the calculation gives a
closed-string setting in which the measure, contour prescription, branch
convention, sector contraction, and finite-cutoff sector value can be specified
explicitly~\cite{ManschotWang2024,WangThesis2025}.

The same modular language becomes genuinely dynamical for non-vacuum amplitudes. A one-loop four-point amplitude contains puncture integrations and a Koba--Nielsen kernel,
\begin{equation}
\cA_4^{(1)}
=
\mathcal R^4
\int_{\cF}\frac{\dd^2\tau}{\tau_2^5}
\int_{(T^2)^3}\prod_i \dd^2 z_i\,
\exp\!\left[
\sum_{i<j}s_{ij}G(z_{ij}|\tau)
\right],
\label{eq:discussion-four-point}
\end{equation}
and its low-energy expansion is organized by modular graph functions and
related non-holomorphic modular objects
~\cite{GreenRussoVanhove2008,DHokerGreenVanhove2015ModularStructure,DHokerGreenGurdoganVanhove2017}.
In that setting the cusp data are tied to genuine threshold structure and
worldsheet unitarity rather than to a final supersymmetric zero.  The vacuum
calculation isolates the corresponding regularization mechanism in the closed
oriented torus setting: the Lorentzian/\(E_s\) prescription defines the sector
functional, and the \(\Gamma(2)\) Rademacher reconstruction supplies a second,
cusp-theoretic description of the holomorphic degeneracies entering that same
functional~\cite{ChengDuncan2012,BacciantiChandraEberhardtHartmanMizera2025}.
In this form the Type IIB torus vacuum connects the long-tube contour
prescription with the Rademacher organization of the underlying cusp data,
while the final GSO projection remains an exact coefficient identity.

\acknowledgments

The author thanks Zhi-Zhen Wang for various insightful discussions.

\setcounter{tocdepth}{1}
\addtocontents{toc}{\protect\setcounter{tocdepth}{1}}
\appendix
\section{Conventions and normalizations}
\label{app:conventions}

The conventions used repeatedly in the main text are collected here.  None of
them is individually exotic, but several of them interact in places where sign
errors, factor-of-two mistakes, or naming drift would otherwise obscure the
logic of the vacuum calculation.  In particular, the distinction between the
closed-string sector coefficient \(64\) and the holomorphic comparison
coefficient \(16\), as well as the sign of the
generalized-exponential-integral tail, depends on keeping the conventions
stable from the outset.

\subsection{Nome, theta functions, and eta function}
\label{appsub:nome}

The torus modulus is
\begin{equation}
\tau=x+\ii y,
\qquad
y>0,
\label{eq:tauconv}
\end{equation}
and the nome is defined by
\begin{equation}
q=\ee^{2\pi \ii\tau}.
\label{eq:qconv}
\end{equation}
This is the convention used throughout the present paper and in the associated
coefficient-generation code. It differs from the alternative mathematical
convention \(q_{\rm math}=\ee^{\pi \ii\tau}\), so all exponentials in the
Fourier factors must be fixed relative to \eqref{eq:qconv}.

The Dedekind eta function is
\begin{equation}
\eta(\tau)=q^{1/24}\prod_{n\ge 1}(1-q^n),
\label{eq:etaconv}
\end{equation}
and the relevant Jacobi theta constant is
\begin{equation}
\vartheta_2(\tau)=2q^{1/8}\prod_{n\ge 1}(1-q^n)(1+q^n)^2.
\label{eq:theta2conv}
\end{equation}
For the \(SO(8)\) character rewriting, we also use \(\vartheta_3\) and
\(\vartheta_4\) in the standard genus-one superstring conventions
~\cite{GreenSchwarzWitten1987,GreenSchwarzWitten1988,Polchinski1998Vol2}. The
Jacobi identity takes the form
\begin{equation}
\vartheta_3(\tau)^4-\vartheta_4(\tau)^4-\vartheta_2(\tau)^4=0.
\label{eq:jacobiconv}
\end{equation}

\subsection{\texorpdfstring{$SO(8)$}{SO(8)} characters and sector blocks}
\label{appsub:characters}

The transverse \(SO(8)\) characters are
\begin{equation}
V_8(\tau)=\frac{\vartheta_3(\tau)^4-\vartheta_4(\tau)^4}{2\eta(\tau)^4},
\qquad
S_8(\tau)=\frac{\vartheta_2(\tau)^4}{2\eta(\tau)^4}.
\label{eq:V8S8conv}
\end{equation}
Equation \eqref{eq:jacobiconv} implies
\begin{equation}
V_8(\tau)=S_8(\tau).
\label{eq:VeqSconv}
\end{equation}
The holomorphic sector block used in the vacuum computation is
\begin{equation}
A_V(\tau)=\frac{V_8(\tau)}{\eta(\tau)^8}
=
\frac{\vartheta_2(\tau)^4}{2\eta(\tau)^{12}},
\label{eq:AVconv}
\end{equation}
and similarly for \(A_S\). Therefore
\begin{equation}
A_V(\tau)=A_S(\tau)=8+128q+1152q^2+\cO(q^3).
\label{eq:Aseriesconv}
\end{equation}
The factor \(1/2\) in \eqref{eq:AVconv} is the reason the constant term of the
closed-string sector block is \(8\), not \(16\).

For comparison, the purely holomorphic quantity
\begin{equation}
\frac{\vartheta_2(\tau)^4}{\eta(\tau)^{12}}
=
16+256q+\cO(q^2)
\label{eq:holomorphiccomparison}
\end{equation}
appears naturally in open or unoriented one-loop normalization comparisons and
serves here only as a convention comparison~\cite{AngelantonjSagnotti2002}. It is not the closed type
IIB left-right sector block used in the vacuum theorem.

The constants \(16\), \(8\), and \(64\) therefore enter at different stages of
the normalization.  The first is the constant term of the holomorphic
comparison series \(\vartheta_2^4/\eta^{12}\).  The second is the constant term
of the closed-string holomorphic sector block
\(A_V=V_8/\eta^8\), after the \(SO(8)\)-character normalization has been
included.  The third is the closed left-right coefficient
\(F_{VV}(0,0)=8\cdot 8\), which is the coefficient inserted into the Type IIB
sector functional.

\subsection{Wedge measure and Fourier factors}
\label{appsub:measure}

With
\begin{equation}
\tau=x+\ii y,
\qquad
\bar\tau=x-\ii y,
\label{eq:taubartau}
\end{equation}
one has
\begin{equation}
\dd\tau=\dd x+\ii\dd y,
\qquad
\dd\bar\tau=\dd x-\ii\dd y,
\label{eq:dtaudbar}
\end{equation}
and therefore
\begin{equation}
\dd\tau\wedge\dd\bar\tau=-2\ii\,\dd x\,\dd y.
\label{eq:wedgeconv}
\end{equation}
This is the source of the common factor \(-2\ii\) in the regularized block.

Using the nome convention \eqref{eq:qconv},
\begin{equation}
q^m\bar q^n
=
\ee^{2\pi \ii (m-n)x}\ee^{-2\pi (m+n)y}.
\label{eq:qmqnconv}
\end{equation}
Hence the horizontal strip projection is
\begin{equation}
\int_{-1/2}^{1/2}\ee^{2\pi \ii (m-n)x}\dd x
=
\delta_{m,n},
\label{eq:deltaconv}
\end{equation}
which is why the cusp tail enters only on diagonal modes.

\subsection{Generalized exponential integral}
\label{appsub:Es}

The present paper uses
\begin{equation}
\Es(z)=\int_1^\infty \ee^{-zt}t^{-s}\dd t,
\qquad
\Re(s)>1.
\label{eq:Esconv}
\end{equation}
For \(z>0\) this is equivalent to the incomplete-gamma representation
\begin{equation}
\Es(z)=z^{s-1}\int_z^\infty \ee^{-u}u^{-s}\dd u,
\label{eq:Esgammaconv}
\end{equation}
which is the form often used in the literature following
Manschot and Wang
~\cite{ManschotWang2024,WangThesis2025}. The constant value is
\begin{equation}
\Es(0)=\frac{1}{s-1},
\qquad
E_6(0)=\frac15.
\label{eq:Es0conv}
\end{equation}

In the vacuum application, the diagonal argument is always \(4\pi m\ge 0\), so
the branch-cut issues relevant to genuine threshold problems do not yet arise.

\subsection{Amplitude convention}
\label{appsub:amplitude}

The regularized mode block is written as
\begin{equation}
L^r_{m,n,s}
=
\int_{\cF_1}\dd\tau\wedge\dd\bar\tau\,y^{-s}q^m\bar q^n
-2\ii\,\delta_{m,n}\Es(4\pi m).
\label{eq:Lrconv}
\end{equation}
Because of the wedge measure \eqref{eq:wedgeconv}, the vacuum block is
typically imaginary. The sector integral is therefore written as
\begin{equation}
I^r_{XY}=\sum_{m,n\ge 0}F_{XY}(m,n)L^r_{m,n,6},
\label{eq:Irconv}
\end{equation}
while the real amplitude convention is
\begin{equation}
\cA_{XY}=\ii I^r_{XY}.
\label{eq:appA-Aconv}
\end{equation}
This convention is what turns the exact zero-mode block
\(
L^r_{0,0,6}=-8\ii/(9\sqrt3)
\)
into the real exact contribution
\(
\cA_{00}=512/(9\sqrt3)
\).

These conventions keep the main text free of repeated local definitions. They
also stabilize the comparisons on which the present analysis depends: the
\(SO(8)\) character normalization, the sign of the cusp tail, the
\(q\)-expansion coefficients, and the distinction between auxiliary sector data
and the signed Type IIB projection.

\section{Contour deformation and cutoff independence}
\label{app:contour-deformation}

The Lorentzian contour at the non-separating cusp and its
generalized-exponential-integral tail are written in the main text at the
canonical cutoff \(T_0=1\). Keeping \(T_0\) explicit makes
the contour dependence visible before the canonical specialization. The discussion
contains four ingredients: the contour deformation in the long-tube variable,
the equivalent \(q\)-plane description near \(q=0\), the non-negative
Lorentzian tail evaluation used in the vacuum proof, and the cancellation of
the splitting-height dependence after the compact part and the tail are
combined.

\subsection{Euclidean segment and Lorentzian tail}
\label{appsub:eucl-lor}

Let \(T_0\ge 1\) be a cutoff height in the cusp. The Euclidean long-tube
variable is
\begin{equation}
y=\operatorname{Im}\tau,
\label{eq:ydefB}
\end{equation}
and the basic Lorentzian continuation replaces the semi-infinite Euclidean tail
by the contour
\begin{equation}
\Gamma(T_0)
=
[1,T_0]\cup\{\,T_0+\ii t_L:\; t_L\ge 0\,\}.
\label{eq:GammaT0B}
\end{equation}
The first segment is the finite Euclidean proper-time interval. The second
segment is the vertical Lorentzian branch selected by the string-theoretic
Feynman prescription~\cite{Witten2013IEpsilon,ManschotWang2024,WangThesis2025}.
In the vacuum problem there are no additional puncture moduli, so the contour
issue is concentrated entirely in this one long-tube variable.

For a diagonal Fourier mode \(m=n\), the cusp factor is
\begin{equation}
q^m\bar q^m=\ee^{-4\pi my}.
\label{eq:diagmodeB}
\end{equation}
Hence the Lorentzian tail takes the form
\begin{equation}
\int_{\Gamma(T_0)}\dd z\, z^{-s}\ee^{-4\pi mz}
=
\int_1^{T_0}\dd y\, y^{-s}\ee^{-4\pi my}
\;+\;
\ii\int_0^\infty \dd t_L\,(T_0+\ii t_L)^{-s}\ee^{-4\pi m(T_0+\ii t_L)}.
\label{eq:GammaSplitB}
\end{equation}
This formula displays two structural features: the intermediate contour
decomposition depends on the height \(T_0\), and the Lorentzian branch is
intrinsically complex even when the Euclidean integrand is real.

\subsection{\texorpdfstring{$q$}{q}-plane picture}
\label{appsub:qplane}

The same contour has a simple interpretation in the degeneration variable
\begin{equation}
q=\ee^{2\pi \ii\tau}=\ee^{2\pi \ii x}\ee^{-2\pi y}.
\label{eq:qdefB}
\end{equation}
Along the Euclidean strip, increasing \(y\) moves \(q\) radially toward the
degeneration point \(q=0\). Along the Lorentzian segment \(y=T_0+\ii t_L\), one
instead obtains
\begin{equation}
q
=
\ee^{2\pi \ii x}\ee^{-2\pi T_0}\ee^{-2\pi \ii t_L}.
\label{eq:qLorB}
\end{equation}
Thus the modulus \(|q|=\ee^{-2\pi T_0}\) is fixed, while the phase winds with
\(t_L\). In the \(q\)-plane, the Lorentzian continuation is therefore a circle
or spiral around \(q=0\) at fixed radius. This is the vacuum analogue of the
general picture emphasized in detailed treatments of the stringy
\(i\varepsilon\) prescription: Euclidean proper time approaches the degeneration
point radially, while the Lorentzian continuation circles that same point in
the complexified variable~\cite{Witten2013IEpsilon,WangThesis2025}.

Equation~\eqref{eq:qLorB} is not a new computational representation. It
identifies the contour with the degeneration geometry of the torus rather than
with an arbitrary split of the real axis. The contour prescribes how the
modular integral approaches the non-separating cusp.

\subsection{Non-negative tail evaluation}
\label{appsub:nonnegative-tail}

The Type IIB vacuum proof only requires the non-negative diagonal modes
\(m\ge0\).  For these modes the vertical Lorentzian tail has an elementary
evaluation in terms of the generalized exponential integral.  Let \(T_0>0\),
\(\Re(s)>1\), and \(m\ge0\).  Then
\begin{equation}
\int_{T_0}^{T_0+\ii\infty}\dd z\,z^{-s}\ee^{-4\pi mz}
=
T_0^{\,1-s}\Es(4\pi mT_0),
\label{eq:appB-tail-lemma}
\end{equation}
with \(\Es(0)=1/(s-1)\).  This is the only contour identity from the
Lorentzian prescription that is needed in the vacuum calculation.

First suppose \(m>0\) and set \(a=4\pi m\).  For \(R>0\) define
\begin{align}
J_\ell(R)&=\int_{T_0}^{T_0+\ii R}\dd z\,z^{-s}\ee^{-az},
\label{eq:appB-Jell}\\
J_r(R)&=\int_{T_0}^{T_0+R}\dd z\,z^{-s}\ee^{-az}.
\label{eq:appB-Jr}
\end{align}
Let \(C_R\) be the quarter-circle arc
\begin{equation}
z(\phi)=T_0+R\ee^{\ii\phi},
\qquad
0\le\phi\le\frac{\pi}{2},
\label{eq:appB-arc}
\end{equation}
oriented from \(T_0+R\) to \(T_0+\ii R\), and write
\begin{equation}
J_\phi(R)=\int_{C_R}\dd z\,z^{-s}\ee^{-az}.
\label{eq:appB-Jphi}
\end{equation}
Choose any branch of \(z^{-s}\) whose cut does not intersect the shifted first
quadrant swept out by these paths.  Since \(T_0>0\), the contour stays away from
the origin and from the negative real axis.  Cauchy's theorem applied to the
positively oriented boundary gives
\begin{equation}
J_r(R)+J_\phi(R)-J_\ell(R)=0,
\qquad
J_\ell(R)=J_r(R)+J_\phi(R).
\label{eq:appB-cauchy}
\end{equation}

It remains to show that the arc contribution vanishes.  On \(C_R\),
\(|\dd z|=R\,\dd\phi\) and
\begin{equation}
|z(\phi)|
=
R\left|\ee^{\ii\phi}+\frac{T_0}{R}\right|.
\label{eq:appB-arc-size}
\end{equation}
For \(R\ge2T_0\), this implies
\begin{equation}
|z(\phi)|^{-\Re(s)}
\le
2^{\Re(s)}R^{-\Re(s)}.
\label{eq:appB-arc-power}
\end{equation}
Moreover,
\begin{equation}
\left|\ee^{-az(\phi)}\right|
=
\ee^{-a(T_0+R\cos\phi)}
\le
\ee^{-aT_0}.
\label{eq:appB-arc-exp}
\end{equation}
Hence
\begin{equation}
|J_\phi(R)|
\le
\frac{\pi}{2}\,2^{\Re(s)}\ee^{-aT_0}R^{1-\Re(s)}
\xrightarrow[R\to\infty]{}0 .
\label{eq:appB-arc-bound}
\end{equation}
Taking \(R\to\infty\) in \eqref{eq:appB-cauchy} gives
\begin{equation}
\int_{T_0}^{T_0+\ii\infty}\dd z\,z^{-s}\ee^{-az}
=
\int_{T_0}^{\infty}\dd y\,y^{-s}\ee^{-ay}.
\label{eq:appB-vertical-real}
\end{equation}
The substitution \(y=T_0t\) then gives
\begin{equation}
\int_{T_0}^{\infty}\dd y\,y^{-s}\ee^{-ay}
=
T_0^{\,1-s}\int_1^\infty\dd t\,t^{-s}\ee^{-aT_0t}
=
T_0^{\,1-s}\Es(aT_0),
\label{eq:appB-Es-positive}
\end{equation}
which proves \eqref{eq:appB-tail-lemma} for \(m>0\).

For \(m=0\) the same contour argument applies with the exponential factor
removed.  The arc estimate becomes
\begin{equation}
|J_\phi(R)|
\le
\frac{\pi}{2}\,2^{\Re(s)}R^{1-\Re(s)}
\xrightarrow[R\to\infty]{}0 ,
\label{eq:appB-zero-arc}
\end{equation}
again because \(\Re(s)>1\).  Therefore
\begin{equation}
\int_{T_0}^{T_0+\ii\infty}\dd z\,z^{-s}
=
\int_{T_0}^{\infty}\dd y\,y^{-s}
=
\frac{T_0^{\,1-s}}{s-1}
=
T_0^{\,1-s}\Es(0).
\label{eq:appB-zero-tail}
\end{equation}
This proves \eqref{eq:appB-tail-lemma} for all non-negative diagonal modes.
The overall factor \(-2\ii\) in the regularized block is not part of this
contour lemma; it enters only when the tail is inserted into the torus measure
\(\dd\tau\wedge\dd\bar\tau=-2\ii\,\dd x\,\dd y\).

\subsection{Cutoff-dependent block}
\label{appsub:T0block}

For \(T_0\ge 1\), define the cutoff keyhole domain
\begin{equation}
\cF_{T_0}
=
\left\{
-\frac12\le x\le \frac12,\quad
\sqrt{1-x^2}\le y\le T_0
\right\},
\label{eq:FT0B}
\end{equation}
and the associated compact contribution
\begin{equation}
K^{(T_0)}_{m,n,s}
=
\int_{\cF_{T_0}}\dd\tau\wedge\dd\bar\tau\,y^{-s}q^m\bar q^n.
\label{eq:KT0B}
\end{equation}
In the corresponding \(E_s\)-regularized prescription, the
\(T_0\)-dependent block is
\begin{equation}
L^r_{m,n,s}(T_0)
=
K^{(T_0)}_{m,n,s}
-2\ii\,\delta_{m,n}\,T_0^{\,1-s}\Es(4\pi mT_0).
\label{eq:LrT0B}
\end{equation}
The present paper uses the canonical choice \(T_0=1\), for which
\begin{equation}
L^r_{m,n,s}(1)
=
\int_{\cF_1}\dd\tau\wedge\dd\bar\tau\,y^{-s}q^m\bar q^n
-2\ii\,\delta_{m,n}\Es(4\pi m).
\label{eq:LrT01B}
\end{equation}
Equation~\eqref{eq:LrT0B} displays the intermediate cutoff dependence before
the cancellation proved below.

\subsection{Cutoff-independence}
\label{appsub:T0independence}

The intermediate pieces in \eqref{eq:LrT0B} depend on \(T_0\), but the full
regularized object does not. The easiest way to see this at the vacuum block
level is to differentiate with respect to \(T_0\). The derivative of the
compact-domain term is the boundary contribution at \(y=T_0\),
\begin{equation}
\frac{\partial}{\partial T_0}K^{(T_0)}_{m,n,s}
=
-2\ii\,\delta_{m,n}\,T_0^{-s}\ee^{-4\pi mT_0},
\label{eq:dKT0}
\end{equation}
where the Kronecker delta comes from the horizontal projection on
\(-\tfrac12\le x\le \tfrac12\).

For the special-function tail, write
\begin{equation}
F(T_0)=T_0^{\,1-s}\Es(4\pi mT_0).
\label{eq:FT0def}
\end{equation}
Using the integral definition
\(
\Es(z)=\int_1^\infty \ee^{-zt}t^{-s}\dd t
\)
and differentiating under the integral sign gives
\begin{equation}
\frac{\dd}{\dd T_0}F(T_0)
=
-T_0^{-s}\ee^{-4\pi mT_0}.
\label{eq:dFT0}
\end{equation}
Therefore
\begin{equation}
\frac{\partial}{\partial T_0}
\Bigl[-2\ii\,\delta_{m,n}F(T_0)\Bigr]
=
2\ii\,\delta_{m,n}\,T_0^{-s}\ee^{-4\pi mT_0},
\label{eq:dTailT0}
\end{equation}
which cancels \eqref{eq:dKT0}. Hence
\begin{equation}
\frac{\partial}{\partial T_0}L^r_{m,n,s}(T_0)=0.
\label{eq:dLrT0zero}
\end{equation}

Thus the cutoff is an intermediate contour parameter. The final block is
independent of the chosen matching height once the compact and tail pieces are
combined.

\subsection{Cusp block in the sector functional}
\label{appsub:cusp-block-sector-functional}

The main text fixes \(T_0=1\), but the preceding derivation shows that this
choice is only a convenient representative of a cutoff-independent block.  The
long-tube contour, its \(q\)-plane description, and the non-negative vertical
tail evaluation \eqref{eq:appB-tail-lemma} define the same cusp contribution
that appears in the \(E_s\)-regularized block.  The \(T_0\)-dependent expression
makes the cancellation between the compact keyhole and the Lorentzian tail
explicit before the value \(T_0=1\) is chosen in the main text.

This is the cusp block that enters the vacuum sector functional.  In non-vacuum
amplitudes the same long-tube data can be embedded into a broader all-cusp
organization.

\section{Constant-mode integral}
\label{app:constant-mode-integral}

The exact zero-mode evaluation used in Section~\ref{sec:constant-mode}
shows how the compact finite part, the constant cusp tail, and the
closed-string coefficient combine inside the same \(E_s\)-regularized
modular-integral prescription used for all modes
~\cite{ManschotWang2024,WangThesis2025}.

\subsection{Zero-mode block as a finite-part functional}
\label{appsub:block}

For \(m=n=0\), the regularized block is
\begin{equation}
L^r_{0,0,6}
=
\int_{\cF_1}\dd\tau\wedge\dd\bar\tau\,y^{-6}
-2\ii E_6(0).
\label{eq:L00}
\end{equation}
Using
\begin{equation}
\dd\tau\wedge\dd\bar\tau=-2\ii\,\dd x\,\dd y,
\label{eq:wedge}
\end{equation}
define the real compact finite part at general \(s\) by
\begin{equation}
I_{\cF_1}^{\rm real}(0,0,s)
=
\int_{-1/2}^{1/2}\dd x
\int_{\sqrt{1-x^2}}^{1}\dd y\,y^{-s}.
\label{eq:Ireal-general-def}
\end{equation}
Then
\begin{equation}
L^r_{0,0,6}
=
-2\ii\Bigl[I_{\cF_1}^{\rm real}(0,0,6)+E_6(0)\Bigr].
\label{eq:L00real}
\end{equation}
The lower boundary \(y=\sqrt{1-x^2}\) is the circular arc of the compact
keyhole, and the upper boundary \(y=1\) is the chosen split between compact
domain and cusp strip.  This is the zero-mode instance of the same compact
part plus diagonal tail structure used in \(L^r_{m,n,s}\).

\subsection{General \texorpdfstring{$s$}{s} formula and \texorpdfstring{$s=6$}{s=6} specialization}
\label{appsub:keyhole}

For \(\Re(s)>1\), the compact finite part is
\begin{equation}
I_{\cF_1}^{\rm real}(0,0,s)
=
\frac{1}{s-1}
\left[
\int_{-1/2}^{1/2}(1-x^2)^{(1-s)/2}\dd x
-1
\right].
\label{eq:Ireal-general}
\end{equation}
Equivalently,
\begin{equation}
I_{\cF_1}^{\rm real}(0,0,s)
=
\frac{2}{s-1}
\int_0^{\pi/6}\cos^{2-s}\theta\,\dd\theta
-
\frac{1}{s-1}.
\label{eq:Ireal-general-theta}
\end{equation}
In the form needed for the vacuum measure, the elementary remaining integral is
\begin{equation}
\int_{-1/2}^{1/2}(1-x^2)^{-5/2}\dd x
=
\frac{20}{9\sqrt3}.
\label{eq:appC-x-integral-value}
\end{equation}
Therefore
\begin{equation}
I_{\cF_1}^{\rm real}(0,0,6)
=
\frac15\left(\frac{20}{9\sqrt3}-1\right).
\label{eq:Iexact}
\end{equation}

\subsection{Cusp tail and assembly of the block}
\label{appsub:cusp}

The constant cusp tail at split height \(Y\ge1\) is
\begin{equation}
T_s(Y)
=
\int_Y^\infty y^{-s}\dd y
=
\frac{Y^{1-s}}{s-1}.
\label{eq:appC-tail-Y}
\end{equation}
The compact-domain contribution and \(T_s(Y)\) have compensating dependence on
the artificial split height \(Y\); moving the boundary only transfers the
constant horizontal mode between the compact part and the cusp tail.  With the
canonical split \(Y=1\),
\begin{equation}
T_s(1)=E_s(0)=\frac{1}{s-1}.
\label{eq:Es0}
\end{equation}
Thus \(E_6(0)=1/5\), and
\begin{equation}
I_{\cF_1}^{\rm real}(0,0,6)+E_6(0)
=
\frac{4}{9\sqrt3}.
\label{eq:bracket}
\end{equation}
Substituting into \eqref{eq:L00real} gives the exact regularized zero-mode
block
\begin{equation}
L^r_{0,0,6}
=
-2\ii\left(\frac{4}{9\sqrt3}\right)
=
-\frac{8\ii}{9\sqrt3}.
\label{eq:L00final}
\end{equation}

\subsection{Closed-sector coefficient insertion}
\label{appsub:sector}

The block-level result is inserted into the closed-string sector functional
through the left-right coefficient
\begin{equation}
F_{VV}(0,0)=a_V(0)^2=8\times8=64.
\label{eq:appC-F00}
\end{equation}
With the amplitude convention \(\cA_{XY}=\ii I^r_{XY}\),
\begin{equation}
\cA_{00}^{XY}
=
\ii F_{XY}(0,0)L^r_{0,0,6}.
\label{eq:appC-A00def}
\end{equation}
Equivalently,
\begin{equation}
\cA_{00}^{XY}
=
2F_{XY}(0,0)
\Bigl[I_{\cF_1}^{\rm real}(0,0,6)+E_6(0)\Bigr].
\label{eq:A00struct}
\end{equation}
Using \eqref{eq:bracket} and \eqref{eq:appC-F00},
\begin{equation}
\cA_{00}^{XY}
=
2\cdot64\cdot\frac{4}{9\sqrt3}
=
\frac{512}{9\sqrt3}.
\label{eq:appC-A00final}
\end{equation}

\subsection{Relation to the sector functional}
\label{appsub:constant-mode-sector-functional}

The value \(\cA_{00}^{XY}=512/(9\sqrt3)\) is an exact auxiliary-sector
normalization datum.  It fixes the wedge measure, the constant cusp tail, and
the closed left-right coefficient in one place.  It is neither the full
auxiliary sector value nor a physical Type IIB vacuum energy.  The full
auxiliary sector value also includes non-constant modes, and the physical
vacuum amplitude is obtained only after the signed GSO coefficient is applied.

\section{Numerical evaluation of sector values}
\label{app:numerical-details}

The finite-cutoff evaluation concerns the sector-resolved Type IIB functional
\(\cA^{(M)}_{XY}\).  The value is attached to one spin-structure sector before
the signed GSO contraction.  The physical Type IIB torus vacuum is obtained
only after the four sector functionals are combined with signs.

The algebraic input is the coefficient identity
\[
F_{\rm GSO}(m,n)=F_{VV}(m,n)-F_{VS}(m,n)-F_{SV}(m,n)+F_{SS}(m,n)=0 .
\]
This identity is imposed at the coefficient level.  The numerical material
below evaluates the common sector functional after the
Lorentzian/\(E_s\) prescription, the closed-sector normalization, and the
Fourier cutoff have been fixed.

\subsection{Sector functional}
\label{appsub:cutoff}

For a Fourier cutoff \(M\), define
\begin{equation}
\cA^{(M)}_{XY}
=
\ii\sum_{m,n=0}^{M}F_{XY}(m,n)L^r_{m,n,6},
\qquad
X,Y\in\{V,S\}.
\label{eq:AMapp}
\end{equation}
The quoted value is obtained at
\begin{equation}
M_{\max}=7.
\label{eq:Mmax}
\end{equation}
The cutoff applies only to the Fourier-mode sum.  The non-compact cusp has
already been treated by the Lorentzian prescription and by the corresponding
generalized exponential-integral tail.  The regularized block is
\begin{equation}
L^r_{m,n,6}
=
K_{m,n}-2\ii\,\delta_{m,n}E_6(4\pi m),
\label{eq:app-num-L-block}
\end{equation}
where the compact-domain contribution is
\begin{equation}
K_{m,n}
=
-2\ii
\int_{-1/2}^{1/2}\dd x
\int_{\sqrt{1-x^2}}^1\dd y\,
y^{-6}
\ee^{2\pi\ii(m-n)x}
\ee^{-2\pi(m+n)y}.
\label{eq:app-num-K}
\end{equation}
Thus the finite calculation concerns the compact contribution \(K_{m,n}\) and
the convergence of the Fourier sum in \eqref{eq:AMapp}.  There is no numerical
integration over an infinite Euclidean strip.

\subsection{Compact-domain reduction}
\label{appsub:compact-reduction}

The compact integral in \eqref{eq:app-num-K} is two-dimensional as written,
but the \(y\)-integral can be carried out analytically.  Put
\[
\lambda_{m,n}=2\pi(m+n),
\qquad
\ell_{m,n}=m-n,
\qquad
y_0(x)=\sqrt{1-x^2}.
\]
For \(m+n>0\),
\begin{equation}
\int_{y_0(x)}^1 y^{-6}\ee^{-\lambda_{m,n}y}\dd y
=
\lambda_{m,n}^{5}
\left[
\Gamma\!\left(-5,\lambda_{m,n}y_0(x)\right)
-
\Gamma\!\left(-5,\lambda_{m,n}\right)
\right].
\label{eq:app-y-reduction-positive}
\end{equation}
The compact contribution is therefore the one-dimensional oscillatory integral
\begin{equation}
\begin{aligned}
K_{m,n}
=
{}&-2\ii
\int_{-1/2}^{1/2}\dd x\,
\ee^{2\pi\ii\ell_{m,n}x}
\lambda_{m,n}^{5}
\left[
\Gamma\!\left(-5,\lambda_{m,n}\sqrt{1-x^2}\right)
-
\Gamma\!\left(-5,\lambda_{m,n}\right)
\right],
\\[-1mm]
&\hspace{7.8cm} m+n>0 .
\end{aligned}
\label{eq:app-K-one-dimensional}
\end{equation}
The zero mode is kept separate.  For \(\lambda_{0,0}=0\), the elementary
antiderivative gives
\begin{equation}
K_{0,0}
=
-2\ii
\int_{-1/2}^{1/2}\dd x\,
\frac{(1-x^2)^{-5/2}-1}{5}.
\label{eq:app-K-zero-mode-reduction}
\end{equation}
This is the compact part used in the exact zero-mode calculation.  The
reduction leaves a finite collection of compact integrals, while the
non-compact cusp contribution remains the analytic \(E_6\)-tail in
\eqref{eq:app-num-L-block}.

\subsection{Coefficient identities and finite evaluation}
\label{appsub:coefficient-identities-numerical-evaluation}

At fixed cutoff \(M\), the coefficients in \eqref{eq:AMapp} are algebraic
Fourier coefficients of the \(SO(8)\) character blocks:
\[
A_V(\tau)=A_S(\tau)=\sum_{m\ge0}d_mq^m,
\qquad
F_{XY}(m,n)=d_md_n.
\]
This equality is exact in the vacuum sector.  It implies
\begin{equation}
\cA^{(M)}_{VV}
=
\cA^{(M)}_{VS}
=
\cA^{(M)}_{SV}
=
\cA^{(M)}_{SS}
\qquad
\text{for every }M.
\label{eq:app-cutoff-sector-equality}
\end{equation}
The signed combination is consequently
\begin{equation}
\cA^{(M)}_{\rm GSO}
=
\cA^{(M)}_{VV}
-\cA^{(M)}_{VS}
-\cA^{(M)}_{SV}
+\cA^{(M)}_{SS}
=0
\label{eq:app-cutoff-gso-zero}
\end{equation}
at every cutoff.  Equation~\eqref{eq:app-cutoff-gso-zero} is the exact
coefficient identity associated with the signed projection.  The finite
computation evaluates the common sector value on the left-hand side of
\eqref{eq:app-cutoff-sector-equality}.

The same coefficient data also admit the \(\Gamma(2)\) cusp reconstruction
described in Section~\ref{subsec:fourierdata}.  In that description the
holomorphic degeneracies \(d_m=[q^m]B_2\) are obtained from the Rademacher
coefficient construction applied to the remaining cusp data of the
\(\Gamma(2)\) theta-block orbit
~\cite{ChengDuncan2012,BacciantiChandraEberhardtHartmanMizera2025}.
Substituting these degeneracies into
\eqref{eq:AMapp} gives the same finite-cutoff sector functional as the product
expansion of \(A\).  This identifies the coefficient input relative to cusp
data before the regularized sector sum is evaluated.

\subsection{Sector values}
\label{appsub:sectors}

At the working cutoff the four sector functionals give the same value,
\begin{equation}
I^r_{VV}=I^r_{VS}=I^r_{SV}=I^r_{SS}
=
-32.44755686128456\,\ii,
\label{eq:Irapp}
\end{equation}
or, with the amplitude convention \(\cA=\ii I^r\),
\begin{equation}
\cA_{VV}=\cA_{VS}=\cA_{SV}=\cA_{SS}=32.44755686128456.
\label{eq:Aapp}
\end{equation}
The signed Type IIB contraction is obtained by applying
\eqref{eq:app-cutoff-gso-zero} to the four sector functionals.

\subsection{Relation to the exact zero mode}
\label{appsub:zero-mode-comparison}

The finite sector value should be separated from the exact constant-mode
normalization derived in Section~\ref{sec:constant-mode} and
Appendix~\ref{app:constant-mode-integral}.  The constant-mode contribution is
\begin{equation}
\cA^{XY}_{00}
=
\frac{512}{9\sqrt3}
=
32.844815313898710\ldots .
\label{eq:app-A00compare}
\end{equation}
This number follows from the closed-sector coefficient \(F_{XY}(0,0)=64\) and
the regularized zero-mode block.  The all-mode sector value
\eqref{eq:Aapp} differs by the finite non-constant-mode contribution
\begin{equation}
\cA_{XY}-\cA^{XY}_{00}
=
-0.397258452614150\ldots .
\label{eq:app-nonconstant-shift}
\end{equation}
Equivalently,
\[
\cA^{XY}_{00}-\cA_{XY}
\approx
3.97\times10^{-1}.
\]
The decimal sector value is therefore not a second evaluation of the zero-mode
term.  It is the finite all-mode value of one sector before the signed GSO
contraction.

\subsection{Cutoff convergence}
\label{appsub:convergence}

Table~\ref{tab:app-convergence} shows the stabilization of the same sector
functional as the Fourier cutoff is raised.  By \(M=4\), the displayed value
has stabilized at the \(10^{-6}\) level, and the final displayed change is of
order \(10^{-12}\).  This is much smaller than the finite non-constant-mode
shift in \eqref{eq:app-nonconstant-shift}.  The table therefore checks the
finite-cutoff evaluation of the sector functional; the GSO-signed column is
zero for the coefficient-level reason in \eqref{eq:app-cutoff-gso-zero}.

\begin{table}[t]
\centering
\begin{tabular}{@{}ccc@{}}
\toprule
\(M\) & \(\operatorname{Im}I^{r,(M)}_{VV}\) & \(|I^{r,(M)}_{\rm GSO}|\) \\
\midrule
1 & \(-32.44227507819203\) & \(0\) \\
2 & \(-32.44763489355489\) & \(0\) \\
3 & \(-32.44755573857176\) & \(0\) \\
4 & \(-32.44755687680042\) & \(0\) \\
5 & \(-32.44755686107898\) & \(0\) \\
6 & \(-32.44755686128721\) & \(0\) \\
7 & \(-32.44755686128456\) & \(0\) \\
\bottomrule
\end{tabular}
\caption{Cutoff convergence of the sector-resolved functional. The first
numerical column gives the \(VV\) sector value. The second column gives the
signed Type IIB contraction determined by the coefficient identity before
finite numerical evaluation.}
\label{tab:app-convergence}
\end{table}

\subsection{Relation to exact identities}
\label{appsub:numerical-relation-to-identities}

The exact identities used in the main text are upstream from the finite-cutoff
material.  The GSO cancellation is the coefficient identity
\[
F_{\rm GSO}(m,n)=0,
\qquad
\cA_{\rm IIB}=0,
\]
and the zero-mode normalization is the closed-form value
\[
\cA^{XY}_{00}=\frac{512}{9\sqrt3}.
\]
Neither identity is inferred from the decimal sector value.

The finite-cutoff value
\[
\cA^{(7)}_{XY}=32.44755686128456
\]
documents the all-mode sector functional associated with the regularized block
after the analytic prescription has been fixed.  The physical Type IIB vacuum
energy is obtained only after the exact signed contraction of the four sectors.

\clearpage
\section{Type IIB sector normalization}
\label{app:typeI-normalization}

The zero-mode normalization in the main text uses the closed-string
coefficient \(F_{XY}(0,0)=64\).  The object-level normalization map relates
this coefficient to the holomorphic constant term \(16\) of the comparison
series
~\cite{AngelantonjSagnotti2002,WangThesis2025}.  The constants \(16\), \(8\),
and \(64\) are not alternative normalizations of one object; they are constant
terms of three different objects.  The Type IIB sector integral is a closed
left-right functional, so its zero-mode insertion uses the two-variable
coefficient \(F_{XY}(0,0)\).

\subsection{Holomorphic and closed-sector normalizations}
\label{appsub:typeI-objects}

The holomorphic comparison series is
\begin{equation}
H(\tau)
=
\frac{\vartheta_2(\tau)^4}{\eta(\tau)^{12}}.
\label{eq:appE-Hdef}
\end{equation}
Using the product formulas
\begin{equation}
\eta(\tau)
=
q^{1/24}\prod_{r\ge1}(1-q^r),
\qquad
\vartheta_2(\tau)
=
2q^{1/8}\prod_{r\ge1}(1-q^r)(1+q^r)^2,
\label{eq:appE-product-formulas}
\end{equation}
one obtains
\begin{equation}
H(\tau)
=
16\prod_{r\ge1}
\frac{(1+q^r)^8}{(1-q^r)^8}
=
16+256q+2304q^2+\cO(q^3).
\label{eq:appE-Hseries}
\end{equation}
The constant \(16\) is therefore the constant term of this one-variable
holomorphic comparison series.

The holomorphic block appearing in the closed Type IIB sector after the
\(SO(8)\)-character normalization is
\begin{equation}
A_V(\tau)=A_S(\tau)
=
\frac{\vartheta_2(\tau)^4}{2\eta(\tau)^{12}}
=
\frac12 H(\tau).
\label{eq:appE-Adef}
\end{equation}
Thus
\begin{equation}
A_V(\tau)=A_S(\tau)
=
8+128q+1152q^2+\cO(q^3).
\label{eq:appE-Aseries}
\end{equation}
The constant \(8\) is the constant term of the closed-sector holomorphic block
\(A_X\), not of the comparison series \(H\).

\subsection{Character normalization and left-right coefficients}
\label{appsub:typeI-projector}

The factor \(1/2\) in \eqref{eq:appE-Adef} is the \(SO(8)\)-character
normalization:
\begin{equation}
S_8(\tau)
=
\frac{\vartheta_2(\tau)^4}{2\eta(\tau)^4}.
\label{eq:appE-S8}
\end{equation}
Dividing by the remaining oscillator factor gives
\begin{equation}
\frac{S_8(\tau)}{\eta(\tau)^8}
=
\frac{\vartheta_2(\tau)^4}{2\eta(\tau)^{12}}.
\label{eq:appE-S8-over-eta8}
\end{equation}
Once the torus integrand has been written in \(SO(8)\)-character language, this
character normalization has already been incorporated into the holomorphic
sector block.

The closed-string sector coefficient is a left-right product.  Write
\begin{equation}
A_X(\tau)=\sum_{m\ge0}a_X(m)q^m,
\qquad
\cZ_{XY}(\tau,\bar\tau)
=
A_X(\tau)\overline{A_Y(\tau)}
=
\sum_{m,n\ge0}F_{XY}(m,n)q^m\bar q^n.
\label{eq:appE-Fdef}
\end{equation}
Then
\begin{equation}
F_{XY}(m,n)=a_X(m)a_Y(n).
\label{eq:appE-F-product}
\end{equation}
For the constant mode,
\begin{equation}
a_V(0)=a_S(0)=8,
\qquad
F_{VV}(0,0)=8\cdot8=64.
\label{eq:appE-F64}
\end{equation}
The sector integral depends on the closed left-right coefficient
\(F_{XY}(m,n)\).  It does not use the one-variable coefficient of \(H\).

\subsection{Zero-mode normalization map}
\label{appsub:typeI-zero-mode-map}

Appendix~\ref{app:constant-mode-integral} gives the block-level zero-mode
value
\begin{equation}
L^r_{0,0,6}
=
-2i\left(I_{\mathcal F_1}^{\rm real}(0,0,6)+E_6(0)\right)
=
-\frac{8i}{9\sqrt3},
\label{eq:appE-Lzero}
\end{equation}
with
\begin{equation}
I_{\mathcal F_1}^{\rm real}(0,0,6)+E_6(0)
=
\frac{4}{9\sqrt3}.
\label{eq:appE-real-plus-tail}
\end{equation}
Inserting the closed left-right coefficient gives
\begin{equation}
\cA_{00}^{XY}
=
2F_{XY}(0,0)
\left(I_{\mathcal F_1}^{\rm real}(0,0,6)+E_6(0)\right).
\label{eq:appE-sector-A00-struct}
\end{equation}
Using \(F_{XY}(0,0)=64\),
\begin{equation}
\cA_{00}^{XY}
=
2\cdot64\cdot\frac{4}{9\sqrt3}
=
\frac{512}{9\sqrt3}.
\label{eq:appE-sector-A00}
\end{equation}

\clearpage
\section{Sector-level Lorentzian--\texorpdfstring{$E_s$}{Es} equivalence}
\label{app:unified-lorentz-es}

The sector comparison is formulated as an identity of three linear functionals
acting on the full non-polar Fourier series of each Type IIB vacuum sector
before the final GSO projection.

\subsection{Closed-torus sector data}
\label{appsub:typeI-IIB-dictionary}

The regularized long-tube functional of
Manschot and Wang~\cite{ManschotWang2024,WangThesis2025} is specialized to the
closed oriented torus.  The integration domain is the modular fundamental
domain, the measure fixes the exponent to \(s=6\), and the sector blocks are
the four closed left-right products
\[
\cZ_{XY}(\tau,\bar\tau)=A_X(\tau)\overline{A_Y(\tau)},
\qquad X,Y\in\{V,S\}.
\]
Their Fourier matrices \(F_{XY}(m,n)\) have non-polar support \(m,n\ge0\).
At the standard cusp, the tail identity is applied to these closed-string
coefficient matrices in both the Lorentzian and \(E_s\) descriptions.  The
signed Type IIB GSO contraction is a subsequent operation on the sector labels
and implements the Jacobi identity pointwise.  Thus the functional comparison
is a statement about the regularized sector blocks before the final signed
contraction.

\subsection{Functional identity}
\label{appsub:functional-identity}

For each \(X,Y\in\{V,S\}\), the sector block has the non-polar
expansion
\begin{equation}
\cZ_{XY}(\tau,\bar\tau)
=
A_X(\tau)\overline{A_Y(\tau)}
=
\sum_{m,n\ge0}F_{XY}(m,n)q^m\bar q^n.
\label{eq:appF-nonpolar}
\end{equation}
The lower bounds \(m,n\ge0\) are not an additional regulator assumption.  They
follow from the theta/eta product expansion established in
Section~\ref{sec:type-iib-torus-amplitude}.  Consequently the ordinary
Euclidean vacuum-sector integral is already convergent at the standard cusp.

Define the \(E_s\)-regularized functional by the block sum used in
Section~\ref{sec:sector-amplitudes},
\begin{equation}
\mathcal R_{E_s}[\cZ_{XY}]
=
\sum_{m,n\ge0}F_{XY}(m,n)L^r_{m,n,6},
\label{eq:appF-REs}
\end{equation}
where \(L^r_{m,n,6}\) is the compact keyhole contribution plus the diagonal
\(E_6(4\pi m)\) tail in \eqref{eq:reg-Lr-six}.  The corresponding Lorentzian
functional is
\begin{equation}
\begin{aligned}
\mathcal R_{\rm Lor}[\cZ_{XY}]
&=
\sum_{m,n\ge0}F_{XY}(m,n)
\int_{\cF_1}\dd\tau\wedge\dd\bar\tau\,
y^{-6}q^m\bar q^n
\\
&\hspace{1.2cm}
-2\ii\sum_{m\ge0}F_{XY}(m,m)
\int_1^{1+\ii\infty}\dd z\,z^{-6}\ee^{-4\pi mz}.
\end{aligned}
\label{eq:appF-RLor}
\end{equation}
The ordinary Euclidean sector functional is
\begin{equation}
\mathcal I_{\rm ord}[\cZ_{XY}]
=
\int_{\cF}\dd\tau\wedge\dd\bar\tau\,
y^{-6}\cZ_{XY}(\tau,\bar\tau),
\label{eq:appF-Iord-def}
\end{equation}
where the integral is convergent for the non-polar sector blocks considered
here.

\paragraph{Sector equivalence.}
For each auxiliary sector-resolved Type IIB torus-vacuum block
\(\cZ_{XY}\), with \(X,Y\in\{V,S\}\), the three sector functionals agree:
\begin{equation}
\mathcal R_{\rm Lor}[\cZ_{XY}]
=
\mathcal R_{E_s}[\cZ_{XY}]
=
\mathcal I_{\rm ord}[\cZ_{XY}].
\label{eq:appF-functional-identity}
\end{equation}

\paragraph{Proof.}
Decompose the fundamental domain into the compact keyhole \(\cF_1\) and the
standard cusp strip \(1\le y<\infty\), \(-1/2\le x\le1/2\).  In the cusp strip,
inserting \eqref{eq:appF-nonpolar} gives
\begin{equation}
\cZ_{XY}(\tau,\bar\tau)
=
\sum_{m,n\ge0}F_{XY}(m,n)
\ee^{2\pi\ii(m-n)x}\ee^{-2\pi(m+n)y}.
\label{eq:appF-strip-series}
\end{equation}
Horizontal integration projects to the diagonal:
\begin{equation}
\int_{-1/2}^{1/2}\dd x\,
\ee^{2\pi\ii(m-n)x}
=
\delta_{m,n}.
\label{eq:appF-strip-projector}
\end{equation}
The cusp contribution to \(\mathcal I_{\rm ord}\) is therefore
\begin{equation}
-2\ii\sum_{m\ge0}F_{XY}(m,m)
\int_1^\infty\dd y\,y^{-6}\ee^{-4\pi my}.
\label{eq:appF-euclidean-cusp}
\end{equation}
By the definition of \(E_s\),
\begin{equation}
\int_1^\infty y^{-6}\ee^{-4\pi my}\dd y=E_6(4\pi m),
\qquad m\ge0.
\label{eq:appF-tail-identity}
\end{equation}
The \(m=0\) term is included in the same formula because
\(E_6(0)=1/5\).  Combining the compact keyhole integral with
\eqref{eq:appF-euclidean-cusp} and \eqref{eq:appF-tail-identity} gives exactly
the \(E_s\)-regularized sum \eqref{eq:appF-REs}.  Hence
\begin{equation}
\mathcal R_{E_s}[\cZ_{XY}]
=
\mathcal I_{\rm ord}[\cZ_{XY}].
\label{eq:appF-REs-Iord}
\end{equation}

It remains to compare the Lorentzian and \(E_s\) tails.  The vertical-tail
lemma proved in Appendix~\ref{app:contour-deformation} gives, at \(T_0=1\),
\begin{equation}
\int_1^{1+\ii\infty}\dd z\,z^{-6}\ee^{-4\pi mz}
=
E_6(4\pi m),
\qquad m\ge0.
\label{eq:appF-lor-tail}
\end{equation}
Substitution of \eqref{eq:appF-lor-tail} into
\eqref{eq:appF-RLor} gives the same compact-plus-diagonal expression as
\eqref{eq:appF-REs}.  Thus
\begin{equation}
\mathcal R_{\rm Lor}[\cZ_{XY}]
=
\mathcal R_{E_s}[\cZ_{XY}].
\label{eq:appF-RLor-REs}
\end{equation}
Equations \eqref{eq:appF-REs-Iord} and \eqref{eq:appF-RLor-REs} give
\eqref{eq:appF-functional-identity}.  The factor \(-2\ii\) in the cusp terms comes
from the wedge measure; it is not part of the one-dimensional contour lemma.

\paragraph{Convergence.}
The exchange of the sector Fourier series with the compact and cusp integrals
is justified within the non-polar vacuum sector.  On \(\cF_1\),
\(y\ge\sqrt{1-x^2}\ge\sqrt3/2\), and therefore
\begin{equation}
|q^m\bar q^n|
\le
\ee^{-\pi\sqrt3(m+n)}.
\label{eq:appF-compact-damping}
\end{equation}
For every \(\beta>0\), Cauchy's estimate applied to the holomorphic block
\(A(\tau)=\sum_{m\ge0}a(m)q^m\) on \(|q|=\ee^{-\beta}\) gives
\begin{equation}
|a(m)|\le C_\beta\,\ee^{\beta m}.
\label{eq:appF-coeff-bound}
\end{equation}
Taking \(\beta<\pi\sqrt3/2\), the compact-domain majorant
\begin{equation}
\sum_{m,n\ge0}
|F_{XY}(m,n)|\,\ee^{-\pi\sqrt3(m+n)}
\label{eq:appF-compact-majorant}
\end{equation}
is absolutely convergent.  For the diagonal cusp tail, the \(m\ge1\) terms obey
\begin{equation}
E_6(4\pi m)
=
\int_1^\infty\ee^{-4\pi mt}t^{-6}\dd t
\le
\frac{\ee^{-4\pi m}}{4\pi m}.
\label{eq:appF-Es-tail-bound}
\end{equation}
Together with \eqref{eq:appF-coeff-bound}, with \(\beta<2\pi\), this gives
\begin{equation}
\sum_{m\ge1}|F_{XY}(m,m)|E_6(4\pi m)<\infty.
\label{eq:appF-tail-convergence}
\end{equation}
The \(m=0\) contribution is finite because \(E_6(0)=1/5\).  These bounds are
sufficient for termwise integration on the compact region and for the diagonal
cusp sum.

\clearpage
\section{Four-graviton threshold comparison}
\label{app:four-graviton-threshold-comparison}

The non-vacuum threshold material collected here is adjacent to the vacuum
calculation. It is logically independent of the sector-functional identity in
Appendix~\ref{app:unified-lorentz-es}: the vacuum result uses only the torus
integrand, the Lorentzian/\(E_s\) mode functional, the convergence and
projector identities, and the coefficient-level GSO identity. The material
identifies the first four-graviton threshold comparison in which the same
Fourier-block language has a nontrivial kinematic target, while keeping that
comparison separate from the vacuum result.

\subsection{First threshold relation}
\label{appsub:first-threshold-bridge}

The simplest non-vacuum comparison is an exact bridge at the first
four-graviton threshold.  Four-point string amplitudes have long provided a
controlled arena for comparing open- and closed-string structures, high-energy
limits, and loop-level integration methods~\cite{KawaiLewellenTye1986,GrossMende1987,GrossMende1988,BernKosower1992LoopAmplitudes}.  The modular-side bare block
is the first genuine \(q^2\) block,
\begin{equation}
q^2=(q_1q_2q_3q_4)^2,
\label{eq:appG-q2}
\end{equation}
whose bare threshold sits at
\begin{equation}
s_\star=4,
\qquad
I_{2222}^{\rm bare}(s,t)
\sim
C_{2222}^{\rm bare}(t)(s-4)^{7/2}.
\label{eq:appG-bare-threshold}
\end{equation}
In the ordered-channel representation, the corresponding channel is
\begin{equation}
(m_D,m_U)=(1,1).
\label{eq:appG-ordered11}
\end{equation}
The extracted threshold channel polynomial is
\begin{equation}
Q_{11}^{\rm thr}(t)
=
36t^2+144t+128.
\label{eq:appG-Q11}
\end{equation}
The Baikov kernel in this channel carries an additional second-order threshold
zero, so the opening is
\begin{equation}
I_{11}^{\rm ord}(s,t)\sim (s-4)^{11/2}.
\label{eq:appG-ordered-opening}
\end{equation}
The threshold bridge relation is
\begin{equation}
I_{11}^{\rm ord}(s,t)
=
d_{11}(t)(s-4)^2 I_{2222}^{\rm bare}(s,t)+\cdots,
\label{eq:appG-bridge}
\end{equation}
with
\begin{equation}
d_{11}(t)
=
-\frac{\pi^2}{16038}
\frac{(9t^2+36t+32)^2(t^2+4t+6)}
{4t^4-99t^3+850t^2-2442t+2248}.
\label{eq:appG-d11}
\end{equation}
This relation is the first non-vacuum threshold datum naturally expressed
in the same Fourier-block language as the vacuum calculation. It is not a
reconstruction of the full four-graviton amplitude.

\subsection{\texorpdfstring{$s=4$}{s=4} threshold hierarchy}
\label{appsub:s4-threshold-channels}

The ordered \((1,1)\) relation must be compared with the light--heavy channels
\((0,4)\) and \((4,0)\), which have the same bare threshold \(s=4\).  The
channel polynomial suppresses these light--heavy contributions at the leading
order relevant for the \((1,1)\) threshold relation. A count based only on
\[
\Delta^{7/2}\Gamma_L\Gamma_R,
\]
would give an apparent \((s-4)^3\) behavior for \((0,4)\). The
Baikov representation contains the squared channel polynomial \(Q^2\).
For
\[
s=4+\epsilon,\qquad
t=t_\star+\epsilon t_\star/4,\qquad
t_L=\epsilon a,\qquad
t_R=\epsilon b,
\]
the exact channel-polynomial extraction gives
\begin{equation}
Q_{04}
=
\epsilon^2 Q_{04}^{(2)}(t_\star,a,b)+O(\epsilon^3),
\label{eq:appG-Q04}
\end{equation}
where
\begin{equation}
Q_{04}^{(2)}
=
\frac32
(t_\star+1)(t_\star+2)(t_\star+3)
(t_\star-4a-4b-8ab).
\label{eq:appG-Q04-leading}
\end{equation}
Thus the light--heavy channel polynomial has at least a quadratic zero
generically.  The resulting threshold ordering is
\begin{equation}
I_{11}=O(\epsilon^{11/2}),\qquad
I_{04}=O(\epsilon^7),\qquad
I_{40}=O(\epsilon^7)
\label{eq:appG-threshold-ordering}
\end{equation}
at generic fixed angle.  At the reference angles \(t_\star=-1,-2,-3\), the factor
\((t_\star+1)(t_\star+2)(t_\star+3)\) vanishes and the light--heavy channels are
further suppressed:
\begin{equation}
I_{04}=O(\epsilon^9),\qquad I_{40}=O(\epsilon^9).
\label{eq:appG-threshold-reference}
\end{equation}
Thus the channels \((0,4)\) and \((4,0)\) do not contribute at the leading
order of the ordered \((1,1)\) first-threshold relation.

\subsection{Data entering the four-graviton threshold}
\label{appsub:four-graviton-threshold-data}

Beyond the first threshold relation, a four-graviton application introduces
additional channel and puncture data.  The first class consists of
channel polynomials: exact low-level \(Q\)-coefficient extractions, their
channelwise cut integrals, and the matching expressions that compare the cut
data with the corresponding diagonal \(E_s\)-branch structure.  The second
class comes from the puncture integrals: puncture-local simplex formulae,
moment and beta-function representations, and chamber-local symbolic
coefficient packages.  The third class concerns the threshold walls themselves,
where near-diagonal, wider off-diagonal, and pair-resolved Fourier
organizations must be glued across the first physical wall at \(s=4\).

These ingredients place cut expressions, puncture-local coefficients, and
Fourier, Laurent, and logarithmic branch data in a common threshold language.
The formulas displayed above are local in the threshold chamber.  A complete
four-point cusp/Fourier generator would additionally require the full
worldsheet kernel and the puncture integrations over all modular images.

\subsection{Separation from the vacuum functional}
\label{appsub:how-to-use-companion}

The vacuum functional depends on the Type IIB torus integrand, the
regularized Lorentzian and \(E_s\) mode blocks, the convergence and projector
identities of Appendix~\ref{app:unified-lorentz-es}, and the coefficient-level
GSO identity.  The four-graviton threshold quantities introduced above belong
to a separate non-vacuum problem.

The threshold calculation is included to identify the nearest non-vacuum
setting in which the same Fourier-block organization acquires kinematic
content.  Once punctures and Koba--Nielsen factors are restored, the cusp data
are tied to physical thresholds rather than to the sector functionals of the
vacuum torus.  The four-graviton formulas in this appendix therefore serve as
a separate threshold comparison, with definitions and consistency conditions
that are independent of the vacuum calculation.

\bibliographystyle{JHEP}
\bibliography{references}

\end{document}